\documentclass[a4paper,11pt]{article}

\usepackage{jheppub} 

\usepackage[T1]{fontenc} 

\usepackage{bm}
\usepackage{latexsym}
\usepackage{dcolumn}
\usepackage{amsmath,amsfonts,amssymb, amsthm}
\usepackage{graphicx,epsfig}
\usepackage{environ}
\usepackage{mathtools}
\usepackage{braket}
\usepackage{float}
\usepackage{mathrsfs}
\usepackage{esint}
\usepackage{amsmath}
\usepackage{tensor}
\usepackage{fancyhdr}
\usepackage{subcaption}
\usepackage{hyperref}
\usepackage{graphicx,epstopdf}
\usepackage{capt-of}

\hypersetup{
	colorlinks   = true, 
	urlcolor     = blue, 
	linkcolor    = blue, 
	citecolor   = red 
}

\def\o{\omega}

\def\no{\nonumber}

\def\d{\delta}

\def\p{\partial}

\def\na{\nabla}

\def\th{\theta}
\def\k{\kappa}
\def\O{\Omega}

\def\t{\tilde}

\def\be{\begin{equation}}
\def\ee{\end{equation}}
\def\ba{\begin{align}}
\def\ea{\end{align}}

\def\mg{\sqrt{-g}}

\def\lm{\mathcal{L}^{(m)}}

\newcommand{\levicivita}{}
\def\levicivita#1#{\tensor#1{\epsilon}}

\title{Thermodynamic structure of a generic null surface and the zeroth law in scalar-tensor theory}
\author[a]{Sumit Dey,}
 \author[b]{Krishnakanta Bhattacharya,}
 \author[a]{and Bibhas Ranjan Majhi}
 \affiliation[a]{Department of Physics, Indian Institute of Technology Guwahati, Guwahati 781039, Assam, India.}
 \affiliation[b]{IUCAA, Post Bag 4, Ganeshkhind,
Pune University Campus, Pune 411 007, India.}
 \emailAdd{dey18@iitg.ac.in}
\emailAdd{krishnakanta@iucaa.in}
\emailAdd{bibhas.majhi@iitg.ac.in}\date{\today}
\abstract{We show that the equation of motion of scalar-tensor theory acquires thermodynamic identity when projected on a generic null surface. The relevant projection is given by $E_{ab}l^ak^b$, where $E_{ab} =8\pi T_{ab}^{(m)}$ represents the equation motion for gravitational field in presence of external matter, $l^a$ is the generator of the null surface and $k^a$ is the corresponding auxiliary null vector. Our analysis is done completely in a covariant way. Therefore all the thermodynamic quantities are in covariant  form and hence can be used for any specific form of metric adapted to a  null surface. We show this both in Einstein and Jordan frames and find that these two frames provide equivalent thermodynamic quantities.  This is consistent with the previous findings for a Killing horizon. Also, a concrete proof of the zeroth law in scalar-tensor theory is provided when the null surface is defined by a Killing vector.
}

\begin{document}
\maketitle
\flushbottom
\noindent

\section{Introduction}
The present understanding of gravitation is premised upon the seminal work of Einstein, who formulated the theory of general relativity (GR) in 1916 which is still hailed as supreme when it comes to describe gravity. Not only the theory is mathematically consistent, but also it has passed all the observational tests, at least within the solar length scale or in the weak gravity range \cite{Dyson:1920cwa, Will:2014kxa}. In spite of all these successes, there are several reasons which imply that the actual behaviour of gravity might deviate significantly from Einstein's GR in the strong gravity region, where GR is not experimentally well-tested. However, in pursuit of developing a more accurate theory of gravity, one has to remember that the Einstein's theory of gravity cannot be ruled out completely because of its sheer success against the observational tests and also due to its infallible predictions: such as the presence of black holes, gravitational waves \textit{etc.}, existence of which were proved later. Therefore, a more accurate version of the theory of gravitation is more likely to be a modified version of Einstein's GR instead of being a radically new theory \cite{Clifton:2011jh,Nojiri:2017ncd}.

The scalar-tensor (ST) theory is one of the most popular among the modified theories of gravity for various reasons \cite{Callan:1985ia, EspositoFarese:2003ze, Elizalde:2004mq, Saridakis:2016ahq, Crisostomi:2016czh, Langlois:2017dyl}. Unlike the Einstein's gravity, the dynamical variable in this theory is not only the second rank symmetric tensor (\textit{i.e.} the metric tensor), but also the scalar degrees of freedom are accounted for this theory in the form of ``non-minimal coupling'' between the scalar field and the curvature. This theory is described in two different frames which raises several issues in the literature, particularly on the equivalence of the physical results described in these two frames \cite{Faraoni:1999hp, ALAL,Faraoni:1998qx, Faraoni:2010yi, Faraoni:2006fx, Saltas:2010ga, Capozziello:2010sc, Padilla:2012ze, Koga:1998un, Jacobson:1993pf, Kang:1996rj, Deser:2006gt, Dehghani:2006xt, Sheykhi:2009vc,Steinwachs:2011zs, Kamenshchik:2014waa, Banerjee:2016lco, Pandey:2016unk, Ruf:2017xon,Karam:2017zno,Bahamonde:2017kbs,Karam:2018squ,Bhattacharya:2017pqc,Bhattacharya:2018xlq,Bhattacharya:2020wdl,Bhattacharya:2020jgk}. The original frame, where the non-minimal coupling is present, is known as the Jordan frame. With the help of a conformal transformation, the non-minimal coupling can be removed and the theory can be expressed equivalently in the Einstein frame. In that case, the curvature and scalar field are separated out and the scalar field behaves like an external source. Now the issues of the two frames are the following: the apparent mathematical equivalence between the two frames via the conformal transformation raises the question on whether the two frames are physically equivalent \cite{Faraoni:1998qx,Faraoni:2010yi,Faraoni:2006fx,Saltas:2010ga,Capozziello:2010sc,Koga:1998un,Kamenshchik:2014waa,Banerjee:2016lco,Pandey:2016unk,Ruf:2017xon,Bhattacharya:2017pqc,Bhattacharya:2018xlq,Bhattacharya:2020wdl,Bhattacharya:2020jgk} or one of the two frames is more physical than the other one \cite{Faraoni:1999hp, ALAL}. The behaviour of energy and other conserved charges under conformal transformations for ST  as well as higher curvature theories of gravity have been studied in \cite{Deser:2006gt}. Here, the authors show that such charges are invariant under conformal transformations provided the conformal factor goes over to unity at infinity. However, a proper thermodynamic description was not developed until the recent works from the present group \cite{Bhattacharya:2017pqc,Bhattacharya:2018xlq,Bhattacharya:2020wdl,Bhattacharya:2020jgk}. In these works \cite{Bhattacharya:2017pqc,Bhattacharya:2018xlq,Bhattacharya:2020wdl,Bhattacharya:2020jgk}, we have shown that the thermodynamic first law can be obtained in the two different frames using the existing well-defined formalisms of general relativity. In addition, our earlier works show that the thermodynamic parameters are exactly equivalent in two different frames. This provides considerable improvement of the previous work \cite{Koga:1998un}, in which the equivalence of the thermodynamic parameters are subject to a few assumptions such as the asymptotic flatness of the spacetime.

The works stated above \cite{Bhattacharya:2017pqc,Bhattacharya:2018xlq,Bhattacharya:2020wdl,Bhattacharya:2020jgk} (describing thermodynamics laws in the two frames of the scalar-tensor theory) are done in the context of the black hole horizon. In Einstein's GR, it is known for a long time that the thermodynamic structure of general relativity is present in any arbitrary null surface \cite{Padmanabhan:2009vy,Parattu:2013gwa,Chakraborty:2015aja,Chakraborty:2015hna,Bhattacharya:2018epn,Maitra:2018saa,Dey:2020tkj} and is not restricted to the black hole horizon. In fact, the thermodynamics of a null surface is very significant in the context of ``Emergent gravity'' paradigm, which was first predicted by Sakharov \cite{Sakharov:1967pk} and later the idea was resurrected by Jacobson by establishing the fact that the Einstein's equation can be obtained as an equation of state from the Clausius relation on a local Rindler horizon \cite{Jacobson:1995ab}. On the other hand, Padmanabhan and his group established the fact that the governing dynamical equations in GR (such as the Einstein's equation) has a thermodynamic structure on the horizon (see the review \cite{Padmanabhan:2009vy}). In particular, we are driven by the fact that the Einstein's equation, when suitably projected on a null surface, takes the form of a thermodynamic identity \cite{Chakraborty:2015aja} (Interestingly, this has been successfully extended to any order Lanczos-Lovelock gravity as well \cite{Chakraborty:2015hna}). Therefore, within ST theory, one needs to check the possibility for developing the first law of thermodynamics for an arbitrary null-surface. This will provide the generality and robustness of the earlier claim on obtaining the thermodynamic structure in this theory. In the process of obtaining the first law for a generic null surface in ST theory, we need to identify certain terms as the temperature to draw the analogy between the gravitational thermodynamics and the conventional thermodynamics. To claim the analogical expression of temperature as the physical thermodynamic quantity, we need to investigate on whether the expression is consistent with other thermodynamic laws, such as the zeroth law. Now, the idea of temperature becomes meaningful only in the equilibrium thermodynamic system; in gravity this is analogous to the Killing horizon \footnote{It is the horizon, which behaves like a thermodynamic object in black hole thermodynamics and the Killing horizon corresponds to a stationary black hole horizon \cite{Hawking:1971vc}.}. So far we know, the zeroth law has not been explored rigorously in ST theory. Therefore, we need to check whether the identified temperature satisfies the zeroth law for the Killing horizon, which is a special category of null-surface and represents the equilibrium thermodynamic system. In summary, the motivation of the present work is straightforward; \textit{i.e.} obtaining the first law for a generic null-surface and proving the zeroth law for the Killing horizon \footnote{Note that the area increase theorem (\textit{i.e.} the second law of black hole thermodynamics) has already been proved in the ST theory \cite{Bhattacharya:2018xlq}.}. Thus, the present work is motivated to fill the gaps in the literature and to establish the thermodynamics of the scalar-tensor gravity in a more concrete manner.

To obtain the thermodynamic laws, we adopt the following method. A null surface is described by a null vector $l^a$, which is the generator of the surface along with a auxiliary null vector $k^a$ (a brief discussion about the null surface has been provided later in the paper). It has been found that the quantity $R_{ab}l^a$ provides several dynamical equations when it is contracted with the normals (\textit{i.e.} $l^b$ and $k^b$) or with projection tensor $q^b_c$. The contraction $R_{ab}l^al^b$ provides the well-known null Raychaudhuri equation \cite{Raychaudhuri:1953yv, Gourgoulhon:2005ng, Poisson:2009pwt}. The NRE has been used in various gravity theories as a crucial input to derive the relevant gravitational field equations emerging from a constitutive relation applied to a local causal horizon \cite{Jacobson:1995ab, Eling:2006aw, Chirco:2009dc, Dey:2017fld}.  Also, the contraction $R_{ab}l^aq^b_c$ provides the Damour-Navier-Stokes equation \cite{DAMOUR,Price:1986yy,Gourgoulhon:2005ng,Padmanabhan:2010rp} (both contractions $R_{ab}l^al^b$ and $R_{ab}l^aq^b_c$ have been studied extensively in the context of ST theory in our earlier work \cite{Bhattacharya:2020wdl}). It has been recently found that the contraction $R_{ab}l^ak^b$ provides the thermodynamic identity for a generic null surface. Initially it was found that when the expression of $R_{ab}l^ak^b$ is expressed in a adapted set of coordinate describing a null-hypersurface (namely the Gaussian null coordinate (GNC)  \cite{Hollands:2006rj,MORALES}), it manifests in the form of the first law of thermodynamics \cite{Chakraborty:2015aja,Chakraborty:2015hna}. Later, it has been proved that $R_{ab}l^ak^b$ can be expressed as a thermodynamic identity in a covariant way \cite{Dey:2020tkj} \textit{i.e.} the choice of any particular coordinate system is not required. The covariant quantities evaluated in GNC reproduces earlier results \cite{Chakraborty:2015aja}.  We adopt this method \cite{Dey:2020tkj} in the context of scalar-tensor gravity and show that the same method works well to obtain the thermodynamic first law in the two frames. In addition, we also prove that the thermodynamic parameters in the two frames are equivalent, as it has been suggested earlier for the stationary black hole horizon (\textit{i.e.} the Killing horizon) \cite{Bhattacharya:2018xlq}. However, as we discuss later, obtaining thermodynamic law in the Jordan frame is quite non-trivial as compared to the Einstein frame, where the latter case is very much similar to that of the Einstein's gravity. Thereafter, we prove the zeroth law for the Killing horizon. To our knowledge, the zeroth law has not been studied extensively for the scalar-tensor theory of gravity. However, there exists some comments in the literature stating that for any sensible definition of zeroth law in ST gravity, the scalar field is required to be constant on the horizon \cite{Faraoni:2010yi}. In our analysis, we show that imposition of such a strong restriction is not required. Instead, what it only requires is that the scalar field needs to be Lie-transported along the direction of the Killing vector \textit{i.e.} the scalar field is required to be independent of only one coordinate, which is along the direction of the generator of the horizon surface.

Let us give an overview of the paper. In Sec. \ref{sectionaction} we begin with a very brief review of the action and field dynamics in the Einstein and Jordan frames. Next we proceed in Sec. \ref{section3} to our essential study of the covariant formulation of the thermodynamic identity established on a generic null hypersurface in the two frames.  This we begin in Sec. \ref{geomnull} by very briefly describing the geometry of the null surface in the two frames. Thereafter we proceed in Sec. \ref{firstlaw} towards our construction of the thermodynamic identity in both the frames. This allows us then to attribute the equivalence of thermodynamic variables in the two frames. Finally, in order to provide a concrete interpretation of the notion of temperature in the two frames, we establish the proof of the zeroth law in Sec. \ref{sectionzeroth}. This proof has been performed in two different ways as applied to Killing horizons in the two frames. In the end, we added five appendices to present details of our calculations.

Before proceeding ahead, we list a word on notations and dimensions. We are working in a spacetime of dimension $d=4$ and have used the metric signature $(-, +, + ,+)$. We use geometrized unit system where $c$, $\hbar$ and $G$ are set to be unity. The lowercsase Latin indices $a$, $b$,... represent the bulk spacetime indices and run from $0$ to $3$. The cordinate indices on our null surface is designated by the Greek symbols $\mu, \nu,...$ and run from $1$ to $3$. The uppercase latin alphabets $A$, $B$,... are reserved for the transverse/angular coordinates of the $2$ dimensional spacelike subspace of our null hypersurface and run from $2$ to $3$.

  \section{Actions and equations of motion in the two frames : A brief review}\label{sectionaction}
 Among the modified theories of gravity, the ST theory is  a much viable and discussed one. In the original {\it Jordan frame}, the scalar field $\phi$ is non-minimally coupled to the Ricci scalar $R$. The total action for the ST theory in the Jordan frame $(\mathcal{M},\boldsymbol{g},\phi)$ is given by,
 \begin{equation}
 \mathcal{A}^{(ST)} = \int_{V} d^{4} x \sqrt{-g}\frac{1}{16 \pi } \Big(\phi R  - \frac{\omega (\phi)}{\phi} g^{ab} \nabla_a \phi \nabla_b \phi - V(\phi)\Big) + \mathcal{A}^{(m)} ~,
 \label{stactionj}
 \end{equation}
 where $\omega (\phi)$ is known as the Brans-Dicke parameter, which is kept as a variable of the scalar-field $\phi$. When $\omega (\phi)$ is considered as the constant parameter, the scalar-tensor theory boils down to the Brans-Dicke theory. Also, $V(\phi)$ corresponds to the arbitrary scalar-field potential and  $\mathcal{A}^{(m)}=\int_V d^4x\sqrt{-g}\lm$ is the ordinary matter action (ordinary in the sense that the matter fields are not coupled to the scalar field $\phi$). 
 The resulting field equation of $g^{ab}$ corresponding to the action (\ref{stactionj}) with a suitable Gibbons-Hawking-York (GHY) surface term is \cite{Bhattacharya:2017pqc, Bhattacharya:2018xlq}
 \begin{equation}
 E_{ab} = \frac{1}{16 \pi} \Big[\phi G_{ab} + \frac{\omega}{2 \phi}g_{ab} \nabla^i \phi \nabla_i \phi - \frac{\omega}{\phi} \nabla_a \phi \nabla_b \phi + \frac{V}{2} g_{ab} - \nabla_a \nabla_b \phi + g_{ab}\nabla_i \nabla^i \phi\Big] = \frac{1}{2} T^{(m)}_{ab} ~,
 \label{feomgabj}
 \end{equation}
 where $T_{ab}^{(m)}=(-2/\mg)\p(\mg\lm)/\p g^{ab}$ represents the matter energy momentum tensor corresponding to $\mathcal{A}^{(m)}$.

 In the Einstein frame we can remove the non-minimal coupling by the following set of conformal transformations on the metric and rescaling of the scalar field respectively,
 \begin{equation}
 \tilde{g}_{ab} = \Omega^2 g_{ab}, 
 \label{gabconformal}
 \end{equation}
 \begin{equation}
 d \tilde{\phi}  = \sqrt{\frac{2\omega(\phi) + 3}{16 \pi}}  \frac{d \phi}{\phi} ~,
 \label{phirescaling}
 \end{equation}
 where $\Omega^2 = \phi$ along with the condition that $\phi > 0$. 
 The related field equation in Einstein frame turns out to be \cite{Bhattacharya:2017pqc, Bhattacharya:2018xlq}
 \begin{equation}
 \tilde{E}_{ab} = \frac{\tilde{G}_{ab}}{16 \pi } - \frac{1}{2} \tilde{\nabla}_a \tilde{\phi} \tilde{\nabla}_b \tilde{\phi} + \frac{1}{4} \tilde{g}_{ab} \tilde{\nabla}^i \tilde{\phi} \tilde{\nabla}_i \tilde{\phi} + \frac{1}{2} \tilde{g}_{ab} U(\tilde{\phi}) = \frac{1}{2} \tilde{T}^{(m)}_{ab} ~,
 \label{feomgabe}
 \end{equation}
 where $U(\tilde{\phi}) = V(\phi)/(16 \pi \phi^2)$
 and $\tilde{T}^{(m)}_{ab} = -\frac{2}{\sqrt{- \t{g}}} \frac{\partial (\sqrt{- \t{g} L^{(m)}})}{\partial \t{g}^{ab}} = \frac{1}{\phi} T_{ab}^{(m)}$ represents the matter energy momentum tensor corresponding to 
 matter action in the Einstein frame. The gravitational field equation in the Einstein frame \eqref{feomgabe}  can be expressed in the similar form of Einstein's equation as $\t G_{ab}=8\pi (\t T_{ab}^{(\t \phi)}+\tilde{T}_{ab}^{(m)})$ where,
 \begin{align}
 \t T_{ab}^{(\t \phi)}=\tilde{\nabla}_a\tilde{\phi}\tilde{\nabla}_b\tilde{\phi}-\frac{1}{2}\tilde{g}_{ab}\tilde{\nabla}^i\tilde{\phi}\tilde{\nabla}_i\tilde{\phi}-\tilde{g}_{ab}U(\tilde{\phi})~.
 \end{align}
  \textit{Throughout this paper we will follow the notation as presented in this section, where the tilde variables are reserved for the Einstein frame and the untilde ones are for the Jordan frame}.
\section{Covariant thermodynamic description on a generic null surface: equivalence between Jordan and Einstein frames}\label{section3}
\subsection{Spacetime foliation of a null-hypersurface}\label{geomnull}
Since the analysis will be done on a generic null surface, it is necessary to introduce the geometry of this surface here. A brief description will be given; details can be followed from \cite{Gourgoulhon:2005ng}.
We consider the $(1+3)$ dimensional spacetime manifold $(\mathcal{M}, g_{ab})$. Therein, lies a generic null hypersurface, which is a three-dimensional sub-manifold, denoted by $\mathcal{H}$ and described by the metric $\gamma_{\alpha\beta}$, which is adapted to the null surface (here, the Greek indices denote the coordinates adapted to the null hypersurface). Since the metric of the null surface is degenerate, there exists vectors $v^{\alpha}$ such that $\gamma_{\alpha\beta}v^{\alpha}=0$ where  $v^{\alpha}$ is defined on the tangent plane of the null surface. We denote the normal to the surface with $l^a$, which satisfies the geodesic condition $l^a \na_al^b=\kappa l^b$ \cite{Gourgoulhon:2005ng} and are the generators of the null surface. Here $\kappa$ denotes the non-affinity parameter of the null geodesics. For a black hole horizon (which is also a null surface), $\kappa$ is identified as the surface gravity of the black hole horizon and is proportional to the Hawking temperature. Since the null surface $\mathcal{H}$ is self-orthogonal, we have $l^a l_a=0$ and one requires another auxiliary null vector $k^a$ to describe the geometry of the null surface. Furthermore, it is considered that the two null-vectors are cross-normalized \textit{i.e.,} $l^ak_a=l_ak^a=-1$. The intersection of the null surface with a $t(x^a) = \text{constant}$ spacelike hypersurface is designated $S_t$. 

The induced metric onto this transverse spacelike $2$-dimensional cross-section $S_t$ in terms of the null vectors is given by, 
\begin{align}
q_{ab}=g_{ab}+l_ak_b+l_bk_a~. \label{QAB2}
\end{align}

With these prerequisites, we now move on to discuss the procedure to obtain the thermodynamic law for a general null hypersurface in the two frames of the scalar-tensor theory.

\subsection{Thermodynamic first law of a generic null surface in scalar-tensor gravity}\label{firstlaw}

As mentioned the introduction, considering the Ricci tensor $R_{ab}$, various components of the vector $R^{a}_{~b} l^b$  provide important dynamical equations in general relativity. 
The component we are interested in i.e. $R_{ab}l^ak^b$, on the null surface for Einstein \cite{Chakraborty:2015aja, Padmanabhan:2002sha, Kothawala:2007em} and Lanczos-Lovelock gravity \cite{Paranjape:2006ca, Kothawala:2009kc, Chakraborty:2015wma} theories yields a  thermodynamic identity which is analogically similar to the first law of conventional thermodynamics. The original discussion was based on a particular form of metric in the vicinity of a null surface written in GNC.  Recently, for this component, a covariant thermodynamic description has been provided in \cite{Dey:2020tkj}.

This covariant description properly reproduces the previous coordinate dependent results in the case of Einstein's gravity. In the present section, we want to check whether this new formulation works well in ST gravity \textit{i.e.}, whether we can obtain similar thermodynamic identity in the both the frames from the recent approach, as prescribed in \cite{Dey:2020tkj}. In this analysis the starting point was a geometric identity
\begin{align}
-\kappa\theta_{({k})}=- D_a\O^a-\O_a\O^a+\theta_{({l})}\theta_{({k})}+l^i\na_i\theta_{({k})}+\frac{1}{2}\  ^{(2)}R-R_{ab} l^a k^b-\frac{1}{2} R~,
\label{RELTNJOR}
\end{align}
where ${{D_a}}$ is the covariant derivative operator defined on the manifold $({S}_t, {q}_{ab})$ and $^{(2)} {R}$
denotes the Ricci scalar associated with the operator ${{D_a}}$.
The above equation can be obtained by taking the trace of the transversal deformation rate equation \cite{Gourgoulhon:2005ng} (the above equation \eqref{RELTNJOR} is also obtained in \cite{Dey:2020tkj}). This identity does not take into account any information of the  dynamics of gravitational field and hence one can use it in any theory of gravity. Below, in this identity the explicit form of field equations for $g_{ab}$ in ST theory will be used in order to investigate our goal.

\subsubsection{Einstein frame}
We start our analysis in Einstein frame as the situation is simpler in this frame. The non-minimal coupling no longer exists in this frame and, the scalar field appears like the external field. In the Einstein frame (${\mathcal{M}}, \tilde{g}_{ab}, \tilde{\phi}$), we assume the existence of a generic null hypersurface $\tilde{\mathcal{H}}$. Let us briefly describe the nature of such a null surface (for details see \cite{Gourgoulhon:2005ng}) in the Einstein frame which is designated by the constant value of the scalar field $\Phi(x^a)$. The null normal $\t{l}_a$ to $\t{\mathcal{H}}$ is given by $\t{l}_a = e^{\t{\rho}} \t{\nabla}_a \Phi$, with $\t{\rho}$ being a scalar function on $\t{\mathcal{H}}$. The integrable null surface $\t{\mathcal{H}}$ is generated by null generators $\t{\boldsymbol{l}}$ satisfying the geodesic equation $\t{l}^a \t{\nabla}_a \t{l}^b = \t{\kappa} \t{l}^b$. The integrability of the null surface is defined by the Frobenius's theorem, which in its dual formulation \cite{Wald:1984rg} reads,
\begin{equation}
	\t{\nabla}_a \t{l}_b - \t{\nabla}_b \t{l}_a = (\t{\nabla}_a \t{\rho}) \t{l}_b - (\t{\nabla}_b \t{\rho}) \t{l}_a ~.
	\label{dualfrobe}
\end{equation} 
 The non affinity parameter of the null generators assumes the value $\t{\kappa} = \t{l^a}\t{\nabla}_a \t{\rho}$. The transverse $2$-dimensional spacelike submanifold of this null hypersurface is designated by $\tilde{S}_t$. In this frame, since every quantity is represented by tilde variable, we express the identity (\ref{RELTNJOR}) in the following form:
\begin{align}
-\t\kappa\t\theta_{(\tilde{k})}=-\t D_a\t\O^a-\t\O_a\t\O^a+\t\theta_{(\tilde{l})}\t\theta_{(\tilde{k})}+\t l^i\t\na_i\t\theta_{(\tilde{k})}+\frac{1}{2}\  ^{(2)}\t R-\t R_{ab}\t l^a\t k^b-\frac{1}{2}\t R~.
\label{RELTN}
\end{align}
From the above equation \eqref{RELTN}, one can obtain the thermodynamic first law considering the virtual displacement along the auxiliary null vector. The idea is the following. We consider the auxiliary null vector field as being parametrized by $\lambda_{(\t k)}$, which means $\t k^i=-dx^i/d\lambda_{(\t k)}$. Here, we put a negative sign in the definition of $\t{\textbf{k}}$ because, in the following, we obtain the change of thermodynamic parameters due to a small virtual displacement along $\t{\textbf{k}}$. Since $\t{\textbf{k}}$ corresponds to the ingoing null vector (which implies $x^i$ decreases with the increase of $\lambda_{(\t k)}$), we need to put additional negative sign so that the change of the thermodynamic parameters remain positive due to the virtual displacement along $\t{\textbf{k}}$. Furthermore, we consider that a set of two null surfaces are located at $\lambda_{(\t k)}=0$ and at $\lambda_{(\t k)}=\d \lambda_{(\t k)}$. A virtual displacement $\d \lambda_{(\t k)}$ implies a shift from one solution of null hypersurface to the other. Then the coordinate variation under the mentioned virtual displacement is given as $\d x^i=-\t k^i \d \lambda_{(\t k)}$. Next, we multiply both sides of the Eq. \eqref{RELTN} with $\d \lambda_{(\t k)}$ (along with an overall factor $1/8\pi$) and integrate it over the two surface $\tilde{\mathcal{S}}_{t}$, which yields
\begin{align}
-\int_{\tilde{S}_t} d^2 x \sqrt{\t q}  ~ \delta \lambda_{(\t k)} \frac{\t \kappa}{2 \pi} \frac{1}{4}\t\theta_{(\tilde{k})}=  \int_{\tilde{S}_t} d^2 x \sqrt{\t q}~\delta \lambda_{(\t k)} \frac{1}{8 \pi}\Big[\frac{1}{2}{^2 \t R} +\t l^i \t  \nabla_i \t\theta_{(\tilde{k})}+\t\theta_{(\tilde{l})}\t\theta_{(\tilde{k})} -\t \Omega_a \t \Omega^a - \t D_A\t \Omega^A\Big] 
\nonumber
\\
- \int_{\tilde{S}_t} d^2 x \sqrt{\t q}~\delta \lambda_{(\t k)}\Big[\t T_{ab}^{(\t\phi)}+\t T_{ab}^{(m)} \Big]\t l^a \t k^b  ~. \label{RELTHER}
\end{align}
In the above we have used the gravitational field equation of the Einstein frame \eqref{feomgabe}. 
The above equation \eqref{RELTHER} can be given the interpretation analogous to the first law of thermodynamics as applied to the null surface via,
\begin{align}
\int_{\tilde{S}_t} d^2 x\t T \delta_{\lambda(\t k)} \t s = \delta_{\lambda(\t k)} \t E + \t F \delta \lambda_{(\t k)} ~,\label{1STLAWEIN}
\end{align}
where, the thermodynamic parameters are identified as the following. We identify the temperature as $\t T=\t\kappa/2\pi$, the entropy density $s$ is identified as $s=\sqrt{\t q}/4$ and, the change of entropy density ($s$) due to the virtual displacement is denoted by $\delta_{\lambda(\t k)} \t s$, which is given as
\begin{align}
\delta_{\lambda(\t k)} \t s =\frac{d\t s}{d \lambda_{(\t k)}} \delta\lambda_{(\t k)} =\frac{1}{4}\frac{d\sqrt{\t q}}{d \lambda_{(\t k)}} \delta\lambda_{(\t k)}=-\frac{1}{4}\sqrt{\t q} ~\t\theta_{(\t{k})}\delta\lambda_{(\t k)}~.\label{CHNGENTROPYDEN}
\end{align}
While obtaining the last step in the above relation \eqref{CHNGENTROPYDEN}, we have used (see \cite{Gourgoulhon:2005ng, Poisson:2009pwt})
\begin{align}
\t\theta_{(\t{k})}=-\frac{1}{\sqrt{\t q}}\frac{d\sqrt{\t q}}{d \lambda_{(\t k)}}~.
\end{align}
The total entropy in Einstein frame is given as
\begin{align}
\t S=\int_{\tilde{S}_t} d^2 x \t s=\frac{1}{4}\int_{\tilde{S}_t} \sqrt{\t q} d^2 x~,
\end{align}
which is consistent with the area law of the entropy.
The variation of energy $\t E$ due to the virtual displacement (in \eqref{1STLAWEIN}) is given as
\begin{align}
\delta_{\lambda(\t k)} \t E =\frac{1}{8 \pi} \int_{\tilde{S}_t} d^2 x \sqrt{\t q} ~ \delta \lambda_{(\t k)}\Big[\frac{1}{2}{^2 \t R} +\t l^i \t  \nabla_i \t\theta_{(\tilde{k})}+\t\theta_{(\tilde{l})}\t\theta_{(\tilde{k})} -\t \Omega_a \t \Omega^a - \t D_A\t \Omega^A\Big] ~.
\label{varEe}
\end{align}
An indefinite integration over $\lambda_{(\t k)}$ provides the expression of energy associated with the two surface $S_t$, which is given as
\begin{align}
\t E =\frac{1}{8 \pi} \int_{\tilde{S}_t} \int d^2 x \sqrt{\t q} ~d \lambda_{(\t k)}\Big[\frac{1}{2}{^2 \t R} +\t l^i \t  \nabla_i \t\theta_{(\tilde{k})}+\t\theta_{(\tilde{l})}\t\theta_{(\tilde{k})} -\t \Omega_a \t \Omega^a - \t D_A\t \Omega^A\Big] ~.
\label{Ee}
\end{align}
The above expression of energy is very much similar to the Hawking-Hayward quasi-local energy \cite{Hayward:1997jp, Prain:2015tda}. The above expression has been identified as the energy term inspired by the fact that it reduces to expressions of that for well known spacetimes. For example, it has been shown in \cite{Dey:2020tkj} that specifically for Einstein gravity the covariant expression of the energy term matches with the expression of the energy expressed in the GNC system \cite{Chakraborty:2015aja}. For example, the covariant energy term for the Schwarzschild metric (in Einstein gravity) reduces to the mass.
Finally, we identify the pressure ($\t P$) as $\t P=-(\t T_{ab}^{(\t\phi)}+\t T_{ab}^{(m)})\t l^a \t k^b$ in the similar way as it has been identified in \cite{Chakraborty:2015aja,Chakraborty:2015wma, Kothawala:2010bf}. Total work due to the virtual displacement $\delta\lambda_{(\t k)}$ is given as  
\begin{align}
\t W=\t F\delta\lambda_{(\t k)}=\int_{\tilde{S}_t} d^2x \sqrt{\t q} ~\delta\lambda_{(\t k)} \t P =-\int_{\tilde{S}_t} d^2x \sqrt{\t q}~\delta\lambda_{(\t k)} (\t T_{ab}^{(\t\phi)}+\t T_{ab}^{(m)})\t l^a \t k^b ~.
\end{align}
Here $\t F$ is the integral of the pressure over the two-surface $S_t$ and hence can be given the interpretation of the generalized force conjugate to the virtual displacement $\delta\lambda_{(\t k)}$. Let us briefly describe the notion of the virtual displacement in the Einstein frame (for details refer to \cite{Chakraborty:2015aja}). In the next subsection, related to Jordan frame, the same interpretation will follow as well. The virtual displacement is considered to be a ``physical process" that virtually shifts the position of the null surface $\t{\mathcal{H}}$ in the Einstein frame from $\lambda_{(\t k)} = 0$ to say $\lambda_{(\t k)} = \delta \lambda_{(\t k)}$. The null hypersurface is obviously considered to be a solution of the field equations in $(\mathcal{M},\t{\boldsymbol{g}}, \t{\phi} )$. As a result of this virtual displacement process, an amount of energy $\delta_{\lambda(\t{k})} \t{E}$ flows across the null hypersurface. Part of this energy contributes in the entropy generation term $\int_{\tilde{S}_t} d^2 x \tilde{T} \delta_{\lambda(\tilde{k})} \tilde{s}$ and the other contributes to the virtual work done $\tilde{F} \delta \lambda_{(\t{k})}$. Let us note, before proceeding next, that all the relevant quantities (geometrical, physical and thermodynamical) in the Jordan frame will be denoted without the use of any tilde as opposed to the Einstein frame.

\subsubsection{Jordan frame}
We now proceed to obtain the thermodynamic law in the Jordan frame taking hints from the analysis in the Einstein frame. Therefore the $R_{ab}l^ak^b$ relation in the Jordan frame is given by (\ref{RELTNJOR}).
We see that the evolution equations for $\tilde{R}_{ab}\tilde{l}^a \tilde{k}^b$ and $R_{ab}l^a k^b$ are form invariant under conformal transformations, viz Eq (\ref{RELTN}) and Eq (\ref{RELTNJOR}). This is to be anticipated since they are evolution equations valid as geometrical identities on any arbitrary null hypersurface. Conformal transformations after all do not alter the causal structure of null hypersurfaces. Infact, it can also be proven that the geodesic equation for the null generators as well as the NRE remains form invariant under conformal transformations for a generic null hypersurface in the Einstein and Jordan frames.
Simply multiplying equation \eqref{RELTNJOR} by $\delta\lambda_{(k)}/8\pi$ and integrating over the two-surface $S_t$ does not lead to the correct expression of thermodynamic law and identification of thermodynamic quantities. The reason is the following. It has been found in the earlier works \cite{Bhattacharya:2017pqc,Bhattacharya:2018xlq,Bhattacharya:2020wdl} for a Killing horizon that the thermodynamic quantities are equivalent in the two frames. Therefore we expect our present thermodynamic quantities, defined on a generic null surface, must be equivalent in the two frames at least when the null normal becomes the symmetry generator of a Killing horizon. Let us now check whether this is the case. If we multiply Eq. (\ref{RELTNJOR}) by $\delta\lambda_{(k)}/8\pi$ and integrate over two surface, the term on left hand side then yields
$-\frac{1}{8\pi}\int_{S_t}d^2x\sqrt{q}~ \delta\lambda_{(k)}\kappa\theta_{(k)} $
which by the earlier argument can be expressed as
\begin{equation}
-\frac{1}{8\pi}\int_{S_t}d^2x\sqrt{q}~\delta\lambda_{(k)}\kappa\theta_{(k)}  = \int_{S_t}d^2x \frac{\kappa}{2\pi}\delta_{\lambda}\Big(\frac{\sqrt{q}}{4}\Big)~.
\label{B1}
\end{equation}
Now if the null surface is a Killing horizon, then $\kappa$ is constant on $S_t$ (we will explicitly prove this later in section \ref{sectionzeroth}). In this case the above is expressed as $(\kappa/2\pi) \delta_\lambda(A/4)$, from which one can identify temperature and entropy as $\kappa/2 \pi$ and $A/4$ respectively. But this is in conflict with the earlier result \cite{Bhattacharya:2017pqc,Bhattacharya:2018xlq} since this is not equivalent to its counter part in Einstein frame. For Killing horizon we know that $\tilde{T} = T$ and $\tilde{S} = \tilde{A}/4 = S=\phi A/4$. But this is not what we are obtaining from the above. Hence the above simple extension of the approach will not be consistent to known cases.

The remedy can be found from the earlier work \cite{Bhattacharya:2020wdl}. From the analysis of fluid-gravity correspondence in scalar-tensor gravity \cite{Bhattacharya:2020wdl}, it is known that the the parameters of the Einstein frame (such as $\t \kappa$, $\t\O^a$, $\t\theta_{(\tilde{l})}$ \textit{etc.}) are used in the Jordan frame as well to obtain the equivalent framework, where the physical parameters of the Einstein frame becomes equivalent to the same in the Jordan frame. 
Here we adopt the same method. Therefore, we plan to obtain the $R_{ab}l^ak^b$ in terms of the parameters of the Einstein frame (such as $\t \kappa$, $\t\O^a$, $\t\theta_{(\t{l})}$, $\t\theta_{(\t{k})}$ \textit{etc.}) and in terms of the the covariant derivative operator and the null vectors of the Jordan frame (\textit{i.e.} $\nabla_i$, $l^a$, $k^a$ \textit{etc.}). The desired relation can be obtained either from \eqref{RELTN} or from \eqref{RELTNJOR} as \eqref{RELTN} and \eqref{RELTNJOR} are equivalent under the conformal transformation (see the Appendix \ref{AppA}). For simplicity, here we obtain it from eq. \eqref{RELTN}. In the Appendix \ref{AppB} we obtain the same equation from eq. \eqref{RELTNJOR}.

Firstly, we show how the different quantities in one frame are connected to the same in the other frame. From the conformal transformation relation \eqref{gabconformal}, we obtain that the null vectors change between the two frames in the following manner \cite{Bhattacharya:2020wdl}
\begin{align}
\t l^a&=l^a, \ \ \ \ \ \ \ \ \ \  \t l_a=\phi l_a &
\no 
\\
\t k^a&=\frac{1}{\phi} k^a, \ \ \ \ \ \  \t k_a= k_a~.& \label{LNK}
\end{align}
Let us take note of the nature of the null hypersurface in the Jordan frame. Its important to stress that we are not considering a different null surface $\mathcal{H}$ in the Jordan frame. The hypersurface is still defined by $\Phi(x^a) = \text{constant}$ in the Jordan spacetime as well. To establish this fact, it is sufficient to prove that $\mathcal{H}$ still represents an integrable hypersurface generated by $\boldsymbol{l}$ under the conformal transformation. Taking cue from \eqref{LNK} and \eqref{dualfrobe}, its quite easy to show that,
\begin{equation}
\t{\nabla}_a \t{l}_b - \t{\nabla}_b \t{l}_a = (\t{\nabla}_a \t{\rho}) \t{l}_b - (\t{\nabla}_b \t{\rho}) \t{l}_a  = \phi (\nabla_a l_b - \nabla_b l_a) + (\nabla_a \phi) l_b - (\nabla_b \phi) l_a	 ~.
\end{equation}
This implies, 
\begin{equation}
(\nabla_a l_b - \nabla_b l_a) = (\partial_a \t{\rho} - \nabla_a \ln\phi) l_b - (\partial_b \t{\rho} - \nabla_a \ln\phi) l_a = (\partial_a \rho) l_b - (\partial_b \rho) l_a ~,
\label{dualfrobj}	
\end{equation}
with the scalar function $\rho$ on $\mathcal{H}$ defined by $\t{\rho} = \rho + \ln \phi + \text{constant}$. The relation \eqref{dualfrobj} guarantees the hypersurface orthogonality of the null surface $\mathcal{H}$ generated by $\boldsymbol{l}$ defined via $l_a = e^{\rho} \nabla_a \Phi$. The non-affinity parameter of the null generators of $\mathcal{H}$ are defined via $\kappa = l^a \nabla_a \rho$.

From the relation \eqref{LNK}, we find,
\begin{align}
\t\th_{(\t{l})}&=\th_{({l})}+l^i\na_i\ln\phi~,
\no 
\\
\t\th_{(\t{k})}&=\frac{1}{\phi}\Big[\th_{({k})}+k^i\na_i\ln\phi\Big],
\no 
\\
\t\k&=\k+l^i\na_i\ln\phi~,
\no 
\\
\t\o_a&=\o_a+\frac{1}{2}\Big[l_ak^i\na_i\ln\phi+\na_a\ln\phi-k_al^i\na_i\ln\phi\Big]~,
\no 
\\
\t\O_a&=\O_a+\frac{1}{2}q^b_a\na_b\ln\phi~,
\label{TRANSREL}
\end{align}

Now, we start from the relation \eqref{RELTN} and change the covariant derivative operator of the Einstein frame (\textit{i.e.} $\t \na_i$) to the covariant derivative of the Jordan frame (\textit{i.e.} $\na_i$). Also, the null vectors of the Einstein frame ($\t l^i$ and $\t k^i$) are transformed to the null vectors of the Jordan frame using eq. \eqref{LNK}. However, we keep other parameters (such as $\t \kappa$, $\t\O^a$, $\t\theta_{(\t{l})}$, $\t\theta_{(\t{k})}$ \textit{etc.}) unchanged. In addition, the Ricci tensor, the intrinsic scalar curvature of the whole manifold and the same of the two-surface are expressed in terms of their Jordan frame's counterpart. With these goals in our mind, we obtain

\begin{align}
\t D_a\t\O^a= D_a\t\O^a+ \t\O^i\na_i(\ln\phi)~. \label{TRAN1}
\end{align}
The intrinsic scalar curvature of the two-surface transforms as \cite{Wald:1984rg}
\begin{align}
^{(2)}\t R=\frac{^{(2)} R}{\phi}-\frac{1}{\phi}D^iD_i(\ln\phi)~. \label{TRAN2}
\end{align}
Also, $\t R_{ab}\t l^a\t k^b+\t R/2$ in eq. \eqref{RELTN} can be identified as $\t G_{ab}\t l^a\t k^b$, which changes under conformal transformation as
\begin{eqnarray}
\t G_{ab}\t l^a\t k^b&=&\frac{G_{ab} l^a k^b}{\phi}+\frac{3}{2\phi}l^ak^b\na_a(\ln\phi)\na_b(\ln\phi)-\frac{l^ak^b}{\phi^2}\na_a\na_b\phi-\frac{1}{\phi^2}\nabla^i\na_i\phi
\nonumber
\\
&+&\frac{3}{4\phi}\na_i(\ln\phi)\na^i(\ln\phi)~. 
\label{TRAN3}
\end{eqnarray}
Using the transformation relations \eqref{TRAN1}, \eqref{TRAN2} and \eqref{TRAN3} in \eqref{RELTN} one obtains the desired $R_{ab}l^ak^b$ relation in the Jordan frame, which is given as
\begin{eqnarray}
&&-\t\kappa\t\theta_{(\t{k})}=- D_a\t\O^a-\t\O^i\na_i(\ln\phi)-\t\O_a\t\O^a+\t\theta_{(\t{l})}\t\theta_{(\t{k})}+ l^i\na_i\t\theta_{(\t{k})}+\frac{1}{2\phi}\  ^{(2)} R-\frac{1}{2\phi}D^iD_i(\ln\phi)
\no 
\\
&&-\Big(\frac{R_{ab} l^a k^b}{\phi}+\frac{R}{2\phi}+\frac{3}{2\phi}l^ak^b\na_a(\ln\phi)\na_b(\ln\phi)-\frac{l^ak^b}{\phi^2}\na_a\na_b\phi-\frac{1}{\phi^2}\nabla^i\na_i\phi \nonumber \\
&&~~~~~+\frac{3}{4\phi}\na_i(\ln\phi)\na^i(\ln\phi)\Big)~. \label{RELATHER}
\end{eqnarray}
To interpret the above relation \eqref{RELATHER} as the thermodynamic identity, we firstly use the field equation in the Jordan frame (\textit{i.e.} eq. \eqref{feomgabj}) in \eqref{RELATHER}, which yields upon multiplication by the scalar field $\phi$ on both sides as,
\begin{eqnarray}
&&-\phi \t\kappa\t\theta_{(\t{k})}=-\phi D_a\t\O^a-\t\O^i\na_i(\ln\phi)- \phi \t\O_a\t\O^a+ \phi \t\theta_{(\t{l})}\t\theta_{(\t{k})}+ \phi l^i\na_i\t\theta_{(\t{k})}+\frac{1}{2}\  ^{(2)} R-\frac{1}{2}D^iD_i(\ln\phi)
\no 
\\
&&-{l^ak^b}\Big[\Big(\frac{2\omega+3}{2}\Big)\Big\{\na_a(\ln\phi)\na_b(\ln\phi)-\frac{1}{2}g_{ab}\na^i(\ln\phi)\na_i(\ln\phi)\Big\}-\frac{V}{2\phi}g_{ab}\Big] \nonumber \\
&&-\frac{8\pi}{\phi}T_{ab}^{(m)}l^ak^b~. \label{RELWITHT}
\end{eqnarray}
The terms inside the square bracket of \eqref{RELWITHT} can be identified as the quantity $8\pi\t T_{ab}^{(\t\phi)}$ as computed in the Jordan frame. Note that the same energy-momentum tensor for $\phi$ field was also obtained in \cite{Bhattacharya:2020wdl} when $R_{ab}l^aq^b_c$ was interpreted as Damour-Navier-Stokes equation in Jordan frame. Also, we know that the energy-momentum tensor of the external matter fields are connected in the two frames as $\t T_{ab}^{(m)}=T_{ab}^{(m)}/\phi$. We now follow the same procedure as that of the Einstein's frame to obtain the first law of thermodynamics. In the Einstein frame we considered the virtual displacement of the null hypersurface from $\lambda_{(\t k)} = 0$ to $\lambda_{(\t k)} = \delta \lambda_{(\t k)}$ \textit{i.e} by an amount of $\delta{\lambda_{(\t k)}}$. We obviously expect this numerical value of the displacement to remain the same when we consider an analogous virtual displacement in the Jordan frame. We have the relation,
\begin{equation}
\delta{x^a} = -\t{k^a} \delta{\lambda_{(\t k)}}  = -\frac{k^a}{\phi} \delta{\lambda_{(\t k)}} = -k^a \delta{\lambda_{k}}~.   
\end{equation}
This above relation allows us to interpret $\delta{\lambda_{(\t k)}} = \phi ~\delta{\lambda_{k}}$. This can also be understood by the following way. We know $\tilde{k}^a = -dx^a/d\lambda_{\tilde{k}}$ and $k^a = -dx^a/d\lambda_{k}$ and as $\tilde{k}^a = k^a/\phi$, we must have $\delta\lambda_{(\tilde{k})} = \phi\delta\lambda_{(k)}$. Hence multiplying the relation \eqref{RELWITHT} with $\delta{\lambda_{(\t k)}}/8 \pi = \phi ~\delta \lambda_{(k)}/8 \pi$ and integrating it over the transverse $2$-surface $S_t$ with the integration measure $\sqrt{q} $, we have,
\begin{eqnarray}
&&-\int_{S_t} d^2 x \phi \sqrt{q} \frac{\t \kappa}{2 \pi} \frac{1}{4}\t\theta_{(\t{k})} ~ \delta \lambda_{(\t{k})}= -\int_{S_t} d^2 x \sqrt{q}\Big[\t T_{ab}^{(\t\phi)}+\frac{ T_{ab}^{(m)}}{\phi} \Big] l^a k^b ~ \delta \lambda_{(\t{k})}
\nonumber
\\
&&+ \int_{S_t} d^2 x\sqrt{q} \frac{\phi}{8 \pi}\Big[\frac{1}{2\phi}{^{(2)}  R} +l^i \nabla_i \t\theta_{(\t{k})}+\t\theta_{(\t{l})}\t\theta_{(\t{k})} -\t \Omega_a \t \Omega^a - \t D_A\t \Omega^A-\t\O^i\na_i(\ln\phi)
\nonumber
\\
&&-\frac{1}{2\phi}D^iD_i(\ln\phi)\Big] ~ \delta \lambda_{(\t{k})}  ~. 
\end{eqnarray}
This allows us to have,
\begin{eqnarray}
&&-\int_{S_t} d^2 x\sqrt{q} ~\delta \lambda_{(k)}\phi^2  \frac{\t \kappa}{2 \pi} \frac{1}{4}\t\theta_{(\t{k})} = -\int_{S_t} d^2 x \sqrt{q}~\delta \lambda_{(k)} \phi\Big[\t T_{ab}^{(\t\phi)}+\frac{ T_{ab}^{(m)}}{\phi} \Big] l^a k^b 
\nonumber
\\
&&+\int_{S_t} d^2 x \sqrt{q}~\delta \lambda_{(k)} \frac{\phi^2}{8 \pi}\Big[\frac{1}{2\phi}{^{(2)}  R} +l^i \nabla_i \t\theta_{(\t{k})}+\t\theta_{(\t{l})}\t\theta_{(\t{k})} -\t \Omega_a \t \Omega^a - \t D_A\t \Omega^A-\t\O^i\na_i(\ln\phi)
\nonumber 
\\
&&-\frac{1}{2\phi}D^iD_i(\ln\phi)\Big]  ~.
\label{RELTHERJOR} 
\end{eqnarray}
As earlier, the above equation \eqref{RELTHERJOR} can be interpreted as the first law of the null-surface in the Jordan frame, which is given as
\begin{align}
\int_{S_t} d^2 x T \delta_{\lambda(k)} s = \delta_{\lambda(k)} E +  F \delta \lambda_{(k)} ~.\label{1STLAWJOR}
\end{align}
For the moment we do not give the  covariantly identified thermodynamical quantities in the Jordan frame. This will be given in the next discussion where we will show their equivalence with those in Einstein frame.

\subsubsection{Thermodynamic equivalence in two frames}
In the following, it will be shown that we, not only obtain the first law of thermodynamics in the two frames, but also the fact that the thermodynamic parameters are equivalent in the two frames. Firstly, we identify the temperature in the Jordan frame as $T=\t\kappa/2\pi$. This is equivalent to the temperature $\t T$ in the Einstein frame. Here, the entropy density ($s$) in the Jordan frame is defined as $s=\sqrt{q}\phi/4$. Therefore,
\begin{align}
\delta_{\lambda(k)} s =\frac{ds}{d\lambda_{(k)}}\delta\lambda_{(k)}=\frac{\delta\lambda_{(k)}}{4}\Big(\phi\frac{d\sqrt{q}}{d\lambda_{(k)}}+\sqrt{q}\frac{d\phi}{d\lambda_{(k)}}\Big)=-\delta\lambda_{(k)}\frac{\sqrt{q}\phi}{4}\Big(\theta_{(k)}+k^i\na_i(\ln\phi)\Big)
\nonumber
\\
=-\frac{1}{4}\phi^2\sqrt{q}\t\theta_{(\t{k})}\delta\lambda_{(k)} = -\frac{1}{4}\sqrt{\t{q}} ~\t{\theta}_{(\t{k})} \delta{\lambda_{(\t{k})}} = \delta_{\lambda(\t{k})} \t{s}~, \ \ \ \ \ \ \ \ \ \ \ \ \ \ \ 
\end{align}
where we have used
\begin{align}
\theta_{(k)}=-\frac{1}{\sqrt{q}}\frac{d\sqrt{q}}{d\lambda_{(k)}}~.
\end{align}
The total entropy in the Jordan frame is defined in the similar way as of the Einstein frame, which is given as
\begin{align}
S=\int_{S_t}sd^2x=\int_{S_t}\phi \frac{\sqrt{q}}{4} d^2x= \int_{\t{S_t}} \frac{\sqrt{\t{q}}}{4} d^2 x = \t S~. \label{ENTJOR}
\end{align}
Here, we have used the fact that $\sqrt{\t q}=\phi\sqrt{q}$. Therefore, we obtain that the entropy density and the entropy in the two frames are equivalent. Also, let us note that the usual area law of entropy is not valid in the Jordan frame. But, the obtained expression of entropy is consistent with earlier observation \cite{Kang:1996rj}.

The variation of the energy in Jordan frame due to the virtual displacement is given as
\begin{align}
\d_{\lambda{(k)}}E=\frac{1}{8 \pi}  \int_{S_t} d^2 x \sqrt{q} ~\delta\lambda_{(k)} \phi^2\Big[\frac{1}{2\phi}{^{(2)}  R} +l^i \nabla_i \t\theta_{(\t{k})}+\t\theta_{(\t{l})}\t\theta_{(\t{k})} -\t \Omega_a \t \Omega^a - \t D_A\t \Omega^A
\nonumber
\\
-\t\O^i\na_i(\ln\phi)-\frac{1}{2\phi}D^iD_i(\ln\phi)\Big] ~.
\label{varEj}
\end{align}
The expression of energy associated with the two-surface $S_t$ is identified as 
\begin{align}
E=\frac{1}{8 \pi} \int_{S_t} \int d^2 x \sqrt{q} ~d\lambda_{(k)} \phi^2\Big[\frac{1}{2\phi}{^{(2)}  R} +l^i \nabla_i \t\theta_{(\t{k})}+\t\theta_{(\t{l})}\t\theta_{(\t{k})} -\t \Omega_a \t \Omega^a - \t D_A\t \Omega^A
\nonumber
\\
-\t\O^i\na_i(\ln\phi)-\frac{1}{2\phi}D^iD_i(\ln\phi)\Big]  ~.
\label{Ej}
\end{align}
Its quite evident using \eqref{TRAN1} and \eqref{TRAN2} that the expresions for the variation of the energy \eqref{varEj} and the energy \eqref{Ej} in the Jordan frame are equivalent to the ones established in the Einstein frame, viz \eqref{varEe} and \eqref{Ee} respectively. Hence we have established the fact that the energy terms are equivalent in both the frames under the process of virtual displacement.

The work done under the virtual displacement process in  the Jordan frame is identified as 
\begin{align}
W=-\int_{S_t} d^2 x ~\sqrt{q} ~\delta\lambda_{(k)}\phi \Big({\t T_{ab}^{(\t\phi)}}+ \frac{T_{ab}^{(m)}}{\phi}\Big) l^a k^b = F \delta \lambda_{(k)}~.
\end{align}
Using the relevant transformations \textit{i.e} $k^a = \phi \t{k}^a$, $\t{T}^{(m)}_{ab} = \frac{1}{\phi}T^{(m)}_{ab}$, $\delta \lambda_{(k)}  = \delta \lambda_{(\t k)}/\phi$ and $\sqrt{\t{q}}= \phi \sqrt{q}$ we obtain,
\begin{eqnarray}
F \delta {\lambda_{(k)}} = -\int_{S_t} d^2 x ~\sqrt{q}~\delta\lambda_{(k)} \phi \Big({\t T_{ab}^{(\t\phi)}}+ \frac{T_{ab}^{(m)}}{\phi}\Big) l^a k^b &=& -\int_{\t{S}_t} d^2x \sqrt{\t{q}} ~\delta\lambda_{(\t{k})} \Big(\t{T}^{(\t{\phi})}_{ab} + \t{T}^{(m)}_{ab}\Big)\t{l}^a \t{k}^b 
\nonumber
\\
&=& \t{F} \delta\lambda_{(\t{k})} ~.
\end{eqnarray}
Hence we see that the work done under the virtual displacement process is equivalent in both the Jordan and the Einstein frames. Even though the work fuction turns out to be equivalent, the pressure terms in the respective frames are not synonymous under our interpretation. We identify the pressure $(P)$ in the Jordan frame as,
\begin{equation}
P = - \Big(\phi \t{T}^{(\t{\phi})}_{ab} + T_{ab}^{(m)}\Big) l^a k^b~.
\end{equation} 
Obviously, the pressure functions in the two frames are not  equivalent \textit{i.e.} $\t P \neq P$. Our identification of the pressure term comes from the fact that the force conjugate to the virtual displacement ${\delta \lambda_{(k)}}$ in the Jordan frame is given as the integral of the pressure term over the $2$-surface $S_t$,
\begin{equation}
F = \int_{S_t} d^2 x \sqrt{q} P~.
\end{equation} 

So far we have seen that, like Einstein's gravity, the ST theory has also similar thermodynamic structure on a generic null surface. We found the thermodynamic quantities on both the frames and constructed them in such a way that they are equivalent. It must be mentioned that this identification of quantities is purely analogy. A comparison with the familiar thermodynamics yields such interpretations. But it may happen that the aforesaid null surface may not be describing an equilibrium system and therefore defining the geometric quantities in terms of thermodynamic entities runs into trouble. Hence the discussion till now has been based on a formal analogy between gravitational equations and conventional thermodynamic identities. On the contrary if the manifold has a Killing horizon present in it (which represents a stationary solution of the gravity theory) then, in the light of constancy of surface gravity on the horizon and area (more generally entropy) increase theorem, the thermodynamic interpretation is much more logical. Having said that we mention that the entropy increase theorem for a Killing horzion in the ST theory has been discussed in literature \cite{Bhattacharya:2018xlq}. But constancy of surface gravity on the equilibrium Killing horizon in this theory, as far we aware of, has not been proven explicitly. Of course, there is a mention in literature that for the zeroth law to hold, the scalar field $\phi$ must be constant on the horizon, i.e. it must not only be independent of the coordinate along the null generator, but also of coordinates on $S_t$. In our point of view the latter restriction is very strong. Therefore we aim to look into this issue here. We will posit the existence of a black-hole spacetime. The Killing vector field is only timelike in some open region of the manifold i.e outside a compact region. We mean that only this open region of the spacetime is stationary. The vanishing norm of the Killing vector field determines the position of the Killing horizon. We will see in the next discussion that existence of a timelike Killing vector field in the stationary region of the spacetime and the scalar field $\phi$ being Lie transported along it are enough to prove the constancy of surface gravity on the horizon. Therefore to obtain the zeroth law in general, $\phi$ can be a function of coordinates on $S_t$.
\section{Study of the zeroth law in both the frames}
\label{sectionzeroth}
Having stated our motivation, we are now going to prove the zeroth law (in other words, constancy of surface gravity) in this section. 
As far as the literature is concerned, the proof of the zeroth law crucially depends on the assumptions in the theory. The assumptions constrain the generality of the proof in turn. As far as we know of, the zeroth law has been proven under three specific assumptions. 
\begin{itemize}
	\item { Use of the gravitational field equations along with the assumption that the non-gravitational and matter fields satisfy the Null Dominant Energy Condition (NDEC): This approach does not assume any extra symmetries of the spacetime other than the existence of a Killing vector field. This has been explicitly proven for the case of Einstein gravity \cite{Wald:1984rg} and Lanczos-Lovelock gravity \cite{Ghosh:2020dkk}. Our proof of the constancy of the surface gravity in this section for the case of ST gravity rests upon this assumption.}
	\item{ Assumption of the existence of bifurcate Killing horizons without the need of any gravitational field equations \cite{Poisson:2009pwt}: This however is restrictive since not all Killing horizons admit a bifurcation $2$-surface.}
	\item{Assumptions of extra symmetries in the spacetime without the need of any field equations: This has been explicitly shown in the case of static and circular (stationary axisymmetric with $t$-$\phi$ reflection symmetry) spacetimes admitting the Killing horizon \cite{Carter1, Racz:1995nh, Straumann:2013spu}. We present a proof (which we hope will add to the existing literature) of the zeroth law for static spacetimes in Appendix \ref{staticzeroth}.}
\end{itemize}
     
Our analysis will be done both in Einstein and Jordan frames. In order to do that we start by constructing the background requisites. 

Let us posit the existence of a Killing vector $\tilde{\boldsymbol{\chi}}$ in the Einstein frame $(\mathcal{M}, \tilde{g}_{ab}, \tilde{\phi})$ defined via,
\begin{equation}
\pounds_{\tilde{\boldsymbol{\chi}}} ~\tilde{g}_{ab} \overset{(\mathcal{M},\tilde{\boldsymbol{g}},\tilde{\phi})}{=} 0 ~.
\label{disc}
\end{equation}  
Using the above fact and $\tilde{g}_{ab} = \phi g_{ab}$, we have,
\begin{equation}
\pounds_{\tilde{\boldsymbol{\chi}}}~g_{ab} \overset{(\mathcal{M}, \bf{g}, \phi)}{=} - \frac{1}{\phi} \Big(\pounds_{\tilde{\boldsymbol{\chi}}}~\phi \Big) g_{ab} ~.
\end{equation}
This shows that provided $\pounds_{\tilde{\boldsymbol{\chi}}}~ \phi  \overset{(\mathcal{M}, \bf{g}, \phi)}{\neq} 0$, the vector field $\tilde{\boldsymbol{\chi}}$ becomes the conformal Killing vector field in the Jordan spacetime $(\mathcal{M}, g_{ab}, \phi)$. However, provided we impose the constraint,
\begin{equation}
\pounds_{\tilde{\boldsymbol{\chi}}}~ \phi  \overset{(\mathcal{M}, \bf{g}, \phi)}{=} 0 ~,
\label{constraint}
\end{equation}
we observe that $\tilde{\boldsymbol{\chi}}$ is also the Killing vector field in the Jordan frame as well. As a matter of field renaming (as per our conventions) we can define the generator of this Killing symmetry in the Jordan spacetime $(\mathcal{M}, g_{ab}, \phi)$ to be $\boldsymbol{\chi}$ and hence $\tilde{\boldsymbol{\chi}}$ and $\boldsymbol{\chi}$ coincide in $(\mathcal{M}, g_{ab}, \phi)$ i.e,
\begin{equation}
\tilde{\chi}^a \overset{(\mathcal{M}, g_{ab}, \phi)}{=} \chi^a ~.
\end{equation}
The above relation has been followed from \cite{Jacobson:1993pf} and has also been imposed in \cite{Koga:1998un}. Obviously, we notice that the contravariant components of the Killing vectors match in the two frames, whereas the covariant vectors are related by the conformal factor.
Hence the constraint (\ref{constraint}) translates to the condition,
\begin{equation}
\chi^a \nabla_a \phi \overset{(\mathcal{M}, g_{ab}, \phi)}{=} 0 ~.
\label{constraintj}
\end{equation} 
We now show what the condition (\ref{constraintj}) implies in the Einstein frame. In fact taking help of the rescaling of the scalar field $\phi$ (\ref{phirescaling}) we can show that,
\begin{equation}
\tilde{\chi}^a \tilde{\nabla}_a \tilde{\phi} = \tilde{\chi}^a \partial_a \tilde{\phi} = \sqrt{\frac{(2 \omega(\phi) + 3)}{16 \pi}} ~\frac{1}{\phi} ~\chi^a \nabla_a \phi ~.
\end{equation}
The above relation clearly implies that setting the constraint $\chi^a \nabla_a \phi \overset{(\mathcal{M}, g_{ab}, \phi)}{=} 0$ in the Jordan frame results in an analogous constraint in the Einstein frame, i.e,
\begin{equation}
\tilde{\chi}^a \tilde{\nabla}_a \tilde{\phi}  \overset{(\mathcal{M},\tilde{g}_{ab}, \tilde{\phi})}{=} 0 ~.
\label{constrainte}
\end{equation}
Having established the connections between the constraints (\ref{constraintj}) and (\ref{constrainte}) in the two frames, we now switch our attention to Killing horizons established in the two respective spacetimes. We reiterate that the Einstein frame $(\mathcal{M},\tilde{g}_{ab}, \tilde{\phi})$ and the Jordan frame $(\mathcal{M}, g_{ab}, \phi)$ admit the Killing vector fields $\tilde{\boldsymbol{\chi}}$ and $\boldsymbol{\chi}$ respectively upon which we have assumed the existence of the constraints (\ref{constraintj}) and (\ref{constrainte}).

A Killing horizon $\tilde{\mathcal{H}}^{(K)}$ in the Einstein frame $(\mathcal{M},\tilde{g}_{ab}, \tilde{\phi})$ admitting the Killing vector field $\tilde{\boldsymbol{\chi}}$ is by definition a null hypersurface of co-dimension one such that $\tilde{\boldsymbol{\chi}}$ is normal to $\tilde{\mathcal{H}}^{(K)}$ and hence coincides with the null generators of $\tilde{\mathcal{H}}^{(K)}$. Under the assumption of the constraint (\ref{constraintj}) and $\phi$ being finite on horizon, we necessarily see that the Killing horizon $\tilde{\mathcal{H}}^{(K)}$ under the conformal transformation (\ref{gabconformal}) and scalar field re-scaling (\ref{phirescaling}) is mapped to a Killing horizon $\mathcal{H}^{(K)}$ in the Jordan frame $(\mathcal{M}, g_{ab}, \phi)$. The null generators of $\mathcal{H}^{(K)}$ coincide with the Killing field $\boldsymbol{\chi}$ on $\mathcal{H}^{(K)}$. {Furthermore, we assume that the respective Killing horizons have (transverse to the null generators) spacelike cross-sections that are closed manifolds}. The null generators satisfy the pregeodesic equation on their respective Killing horizons,
\begin{equation}
\tilde{\chi}^b \tilde{\nabla}_b \tilde{\chi}^a \overset{\tilde{\mathcal{H}}^{(K)}}{=} \tilde{\kappa} \tilde{\chi}^a
\label{geodesice}
\end{equation}
and
\begin{equation}
\chi^b \nabla_b \chi^a \overset{\mathcal{H}^{(K)}}{=} \kappa \chi^a ~,
\label{geodesicj}
\end{equation}
where $\tilde{\kappa}$ and $\kappa$ are the  non-affinity parameters and the surface gravities associated with the null generators $\tilde{\boldsymbol{\chi}}$ and ${\boldsymbol{\chi}}$ of $\tilde{\mathcal{H}}^{(K)}$ and $\mathcal{H}^{(K)}$ respectively. It is worth noticing from \eqref{TRANSREL} and under the constraint \eqref{constraint} imposed on the scalar field that  $\t{\kappa}$ and ${\kappa}$ are same. 

We now shift our attention towards the consideration of the zeroth law of black hole mechanics as applied to the Killing horizons in the two frames. Our proof towards the constancy of the surface gravity in the Killing horizon will demand the dynamical content of the theory, in the sense that we will use the gravitational field equations. The dynamical field equations come into play provided we use some energy conditions. For our case, we will assume that the NDEC holds. We will prove the zeroth law as applied to the Killing horizons in both the frames in two different ways.

\subsection{Approach I}
For the first approach we basically follow \cite{Wald:1984rg}. We observe that the relations (\ref{geodesice}) and (\ref{geodesicj}) are applicable only on the respective Killing horizons. Hence directly applying the derivative operator $\nabla_a$ onto such relations that are only valid on the Killing horizon leads us to problems. In order to prove the constancy of the surface gravity we basically need to take its directional derivative along a vector/tensor field that lies in the tangent plane of the Killing horizon. The Killing horizon being a null surface, makes it impossible to have a well defined projection tensor onto it using only the metric and the null normal. However we can look at the tensor field $\epsilon^{abcd} \chi_a$ which is tangent to the Killing horizon as evident from the fact that $\epsilon^{abcd} \chi_a \chi_b = 0$. Here $\epsilon_{abcd}$ is the spacetime volume form. Hence we can apply the derivative operator $\epsilon^{abcd} \chi_a\nabla_b$ as applied to relations that are valid only on the Killing horizon. Taking the completely antisymmetric nature of the volume form, we may as well take the derivative operator $\chi_{[a} \nabla_{b]}$ as applied to relations valid only on the Killing horizon. For any Killing horizon generated by $\boldsymbol{\chi}$, we have the relation \cite{Wald:1984rg} ,
\begin{equation}
\chi_{[a} \nabla_{b]} \kappa \overset{\mathcal{H}^{(K)}}{=} - \chi_{[a} R_{b]}^{~~f} \chi_f ~,
\label{waldonkilling}
\end{equation}
where $R_{ab}$ stands for the Ricci tensor. 

 \subsubsection{Einstein frame}
Let us now begin our analysis in the Einstein frame $(\mathcal{M},\tilde{g}_{ab}, \tilde{\phi})$ as applied to the Killing horizon $\tilde{\mathcal{H}}^{(K)}$ generated by $\tilde{\boldsymbol{\chi}}$. The resulting equation concerning the change of the surface gravity $\tilde{\kappa}$ along any direction tangent to the Killing horizon $\mathcal{H}^{(K)}$is given by,
\begin{equation}
\tilde{\chi}_{[a} \tilde{\nabla}_{b]} \tilde{\kappa}  \overset{\tilde{\mathcal{H}}^{(K)}}{=} - \tilde{\chi}_{[a} \tilde{R}_{b]}^{~~f} \tilde{\chi}_f ~.
\label{waldonkillinge}
\end{equation}
Using the field equations (\ref{feomgabe}), we compute the R.H.S of (\ref{waldonkillinge}) which leads us to,
\begin{eqnarray}
\tilde{\chi}_{[a} \tilde{R}_{b]}^{~~f} \tilde{\chi}_f &= 16 \pi \Big[\frac{1}{2} \tilde{\chi}_{[a}\tilde{T}^{(m) f}_{~b]} \tilde{\chi}_f + \frac{1}{2} \tilde{\chi}_{[a} \tilde{\nabla}_{b]} \tilde{\phi}~(\tilde{\nabla}^{f} \tilde{\phi} \tilde{\chi}_f) -\frac{1}{4} \tilde{\chi}_{[a} \delta_{b]}^{~f} \tilde{\chi}_f ~(\tilde{\nabla}^i \tilde{\phi} \tilde{\nabla}_i\tilde{\phi}) \nonumber
\\
&-\frac{1}{2}\tilde{\chi}_{[a} \delta_{b]}^{~f} \tilde{\chi}_f ~U(\tilde{\phi}) + \frac{1}{32 \pi} \tilde{\chi}_{[a} \delta_{b]}^{~f} \tilde{\chi}_f ~ \tilde{R}\Big] ~.
\end{eqnarray}
Using the constraint (\ref{constrainte}) as applied to the Einstein frame the above relation simplifies which allows us to express (\ref{waldonkillinge}) as,
\begin{equation}
\tilde{\chi}_{[a} \tilde{\nabla}_{b]} \tilde{\kappa} \overset{\tilde{\mathcal{H}}^{(K)}}{=} - \tilde{\chi}_{[a} \tilde{R}_{b]}^{~~f} \tilde{\chi}_f \overset{\tilde{\mathcal{H}}^{(K)}}{=} -8 \pi \tilde{\chi}_{[a} \tilde{T}^{(m) f}_{b]} \tilde{\chi}_f ~.
\label{zeroth1e}
\end{equation}
Next, we notice that projecting the field equations (\ref{feomgabe}) along the null  generators of $\tilde{\mathcal{H}}^{(K)}$ we have,
\begin{eqnarray}
\tilde{E}_{ab} \tilde{\chi}^a \tilde{\chi}^b &= \frac{1}{16 \pi} \tilde{G}_{ab}\tilde{\chi}^a \tilde{\chi}^b - \frac{1}{2} (\tilde{\chi}^a \tilde{\nabla}_a \tilde{\phi})~ (\tilde{\chi}^b \tilde{\nabla}_b \tilde{\phi}) + \frac{1}{4} \tilde{\boldsymbol{\chi}}^2 ~\tilde{\nabla}^i \tilde{\phi} \tilde{\nabla}_i \tilde{\phi} + \frac{1}{2} \tilde{\boldsymbol{\chi}}^2 U(\tilde{\phi}) ~, \nonumber
\\
&= \frac{1}{2} \tilde{T}^{(m)}_{~ab} \tilde{\chi}^a \tilde{\chi}^b 
\end{eqnarray}
where $\tilde{\boldsymbol{\chi}}^2$ stands for $\tilde{g}_{ab}\tilde{\chi}^a\tilde{\chi}^b$.
Employing the constraint as applied in the Einstein frame (\ref{constrainte}) and the fact that $\tilde{\boldsymbol{\chi}}$ is null on the Killing horizon $\tilde{\mathcal{H}}^{(K)}$, we obtain,
\begin{eqnarray}
\tilde{R}_{ab} \tilde{\chi}^a \tilde{\chi}^b \overset{\tilde{\mathcal{H}}^{(K)}}{=} 8 \pi \tilde{T}^{(m)}_{~ab} \tilde{\chi}^a \tilde{\chi}^b ~.
\label{nec1e}
\end{eqnarray}
As mentioned earlier, we assume that our Killing horizon $\tilde{\mathcal{H}}^{(K)}$ is a null hypersurface provided with the topology $\tilde{\mathcal{H}}^{(K)} \simeq \mathbb{R} \times \mathcal{\t{J}}$, where the spacelike cross-section $\tilde{\mathcal{J}}$ is a closed $2$ dimensional manifold (this is similar to the $S_t$ describing the cross-section of our earlier generic null surface). The induced metric onto the cross-section $\tilde{\mathcal{J}}$ is designated as $\tilde{q}_{ab}$. The null generator $\tilde{\boldsymbol{\chi}}$ satisfies \eqref{disc},
which implies that $\tilde{\boldsymbol{\chi}}$ is a symmetry generator of $\tilde{\mathcal{H}}^{(K)}$. Now since $\tilde{q}_{ab}$ is the metric induced by $\tilde{g}_{ab}$ on $\tilde{\mathcal{J}}$ and the fact that the basis vectors on $\tilde{\mathcal{J}}$ are Lie transported along the null generators, we have the  deformation rate tensor $\tilde{\Theta}_{ab}$ of $\tilde{\mathcal{J}}$ (which coincides with the second fundamental form for an integrable null hypersurface in the absence of torsion) \cite{Gourgoulhon:2005ng} vanishing identically,
\begin{equation}
\tilde{\Theta}_{ab} = \frac{1}{2} \tilde{q}^{c}_{~a} \tilde{q}^{d}_{~b} ~ \pounds_{\tilde{\boldsymbol{\chi}}} \t q_{cd} \overset{\tilde{\mathcal{H}}^{(K)}}{=} 0 ~.
\end{equation}
The irreducible decomposition of the deformation tensor
\begin{equation}
\tilde{\Theta}_{ab} = \frac{1}{2} \tilde{q}_{ab} ~\tilde{\theta} 
_{(\tilde{\boldsymbol{\chi}})} + \tilde{\sigma}_{ab}  ~,
\end{equation}
where $\tilde{\theta} 
_{(\tilde{\boldsymbol{\chi}})}$ denotes the expansion scalar corresponding to the null generator $\tilde{\boldsymbol{\chi}}$ and $\tilde{\sigma}_{ab}$ the shear tensor necessiates the fact that,
\begin{equation}
\tilde{\theta}_{(\tilde{\boldsymbol{\chi}})} \overset{\tilde{\mathcal{H}}^{(K)}}{=} 0 ~~~~~\text{and}~~~~~~ \tilde{\sigma}_{ab}\overset{\tilde{\mathcal{H}}^{(K)}}{=} 0 ~.
\end{equation}
This is precisely because the cross section $\tilde{\mathcal{J}}$ is spacelike in nature. Now we can use the NRE as applied on $\tilde{\mathcal{H}}^{(K)}$ to find the value of $\tilde{R}_{ab}\tilde{\chi}^a\tilde{\chi}^b$. The NRE reads as,
\begin{equation}
\tilde{\chi}^a\tilde{\nabla}_a \tilde{\theta}_{\tilde{\boldsymbol{(\chi)}}} - \tilde{\kappa} \tilde{\theta}_{\tilde{\boldsymbol{(\chi)}}} + \frac{1}{2} \tilde{\theta}^2_{\tilde{\boldsymbol{(\chi)}}} + \tilde{\sigma}_{ab} \tilde{\sigma}^{ab} ~\overset{\tilde{\mathcal{H}}^{(K)}}{=}~ - \tilde{R}_{ab}\tilde{\chi}^a\tilde{\chi}^b ~.
\label{NREe}
\end{equation}
As applied to the specific Killing Horizon $\tilde{\mathcal{H}}^{(K)}$, where we established that the expansion scalar and the shear tensor for $\tilde{\boldsymbol{\chi}}$ vanish, the NRE implies,
\begin{equation}
\tilde{R}_{ab}\tilde{\chi}^a\tilde{\chi}^b ~\overset{\tilde{\mathcal{H}}^{(K)}}{=}~ 0 ~.
\label{Rabchiachibe}
\end{equation}
This entails the fact from (\ref{nec1e}) that,
\begin{equation}
\tilde{T}^{(m)}_{~ab}\tilde{\chi}^a\tilde{\chi}^b ~\overset{\tilde{\mathcal{H}}^{(K)}}{=}~ 0 ~.
\label{nece}
\end{equation}
From the above relation we can conclude that the vector field $\tilde{T}^{(m) a}_{~~~~~~b} \tilde{~\chi}^b$ lies on the tangent plane of the Killing horizon  and hence is either null (collinear to $\t{\boldsymbol{\chi}}$) or spacelike (in the tangent plane of $\t{\mathcal{J}}$). To proceed ahead, we will make the assumption that the matter and the non-gravitational fields in $(\mathcal{M},\tilde{g}_{ab}, \tilde{\phi})$ satisfy the Null Dominant Energy Condition (NDEC). The NDEC states that the vector field $\tilde{W}^a$ defined as,
\begin{equation}
\tilde{W}^a \equiv -\tilde{T}^{(m)a}_{~~~~~b} \tilde{\chi}^b ~
\label{NDEC}
\end{equation}
is future directed causal (null or timelike) for the future directed null generator $\tilde{\boldsymbol{\chi}}$ of $\tilde{\mathcal{H}}^{(K)}$. However, we have already shown that on $\tilde{\mathcal{H}}^{(K)}$, the vector field $\tilde{T}^{(m) a}_{~~~~~~b} \tilde{~\chi}^b$ can either be null or spacelike. Hence the NDEC forces $\tilde{T}^{(m) a}_{~~~~~~b} \tilde{~\chi}^b$ to be null on the Killing horizon and hence collinear to the null generators,
\begin{equation}
-\tilde{T}^{(m) a}_{~~~~~~b} \tilde{~\chi}^b ~\overset{\tilde{\mathcal{H}}^{(K)}}{=}~ \tilde{\alpha} \tilde{\chi}^a ~,
\label{NDECconsequencee}
\end{equation} 
where $\tilde{\alpha}$ is some proprotionality factor. Using the above relation in (\ref{zeroth1e}), we finally end up with the establishment of the zeroth law as applied on the Killing horizon in $(\mathcal{M},\tilde{g}_{ab}, \tilde{\phi})$,
\begin{equation}
\tilde{\chi}_{[a} \tilde{\nabla}_{b]} \tilde{\kappa} \overset{\tilde{\mathcal{H}}^{(K)}}{=} 0 ~.
\label{zerothlawe}
\end{equation}
This basically shows the constancy of the surface gravity over the entire Killing horizon $\tilde{\mathcal{H}}^{(K)}$ established in the Einstein frame $(\mathcal{M},\tilde{g}_{ab}, \tilde{\phi})$ by the null generators $\tilde{\boldsymbol{\chi}}$.

\subsubsection{Jordan frame}
Now, we proceed towards the establishment of the zeroth law in the Jordan frame $(\mathcal{M}, g_{ab}, \phi)$. As again, we reiterate that under the constraint (\ref{constraintj}), the Einstein frame $(\mathcal{M},\tilde{g}_{ab}, \tilde{\phi})$ with the  Killing vector $(\tilde{{\boldsymbol{\chi}}})$ is mapped (under the conformal transformation of the metric and the scaling of the scalar field) to the Jordan frame $(\mathcal{M},{g}_{ab}, {\phi})$ with the  Killing vector field $\boldsymbol{\chi}$. We further posit the existence of a Killing horizon $\mathcal{H}^{(K)}$ in the Jordan frame  where its null generators $\boldsymbol{l}$ coincide with the Killing vector $\boldsymbol{\chi}$,
\begin{equation}
\boldsymbol{l}  ~\overset{\mathcal{H}^{(K)}}{=}~ \boldsymbol{\chi} ~.
\end{equation}
The topology of the Killing horizon in the Jordan frame should remain the same, in the sense that the spacelike cross-section of the null surface is assumed to be a closed manifold. The same analysis towards the fact that the Killing horizon $\mathcal{H}^{(K)}$ is a non-expanding horizon follows. The null Killing vector $\boldsymbol{\chi}$ is a symmetry generator of the Killing Horizon $\mathcal{H}^{(K)}$,
\begin{equation}
\pounds_{\boldsymbol{\chi}} g_{ab} ~\overset{\mathcal{H}^{(K)}}{=}~ 0 ~.
\end{equation}
This again implies that the deformation rate tensor and the second fundamental tensor corresponding to $\mathcal{H}^{(K)}$ vanishes. So does the expansion scalar and the shear tensor corresponding to the null generator $\boldsymbol{\chi}$. Again, application of the NRE for $\boldsymbol{\chi}$, leads us to the fact that,
\begin{equation}
R_{ab} \chi^a \chi^b ~\overset{\mathcal{H}^{(K)}}{=}~ 0 ~.
\label{Rabchiachibj}
\end{equation}
As applied to the Killing horizon $\mathcal{H}^{(K)}$, analogous relation holds regarding the directional derivative of the surface gravity along any vector field tangent to the null surface (\ref{waldonkilling}). Its is quite easy to verify that the R.H.S of (\ref{waldonkilling}) upon application of the field equations in the Jordan frame leads us to,
\begin{eqnarray}
-\chi_{[a} R_{b]}^{~~f} \chi_f &~\overset{(\mathcal{M}, g_{ab}, \phi)}{=}~ -\frac{1}{\phi} \Big( 8 \pi \chi_{[a} T_{b]}^{(m)f} \chi_f - \frac{\omega}{2 \phi} \chi_{[a} \delta_{b]}^{~f}\chi_f~(\nabla_i \phi \nabla^i \phi) + \frac{\omega}{\phi} \chi_{[a} \nabla_{b]} \phi ~ (\chi^f \nabla_f \phi) \nonumber 
\\
& - \frac{1}{2} \chi_{[a} \delta_{b]}^{~f} \chi_f ~ V(\phi) + (\chi_{[a} \nabla_{b]}\nabla^{f} \phi) \chi_f - \chi_{[a} \delta_{b]}^{~f} \chi_f (\nabla_i \nabla^i \phi)\Big) \nonumber
\\
& - \frac{1}{2} \chi_{[a} \delta_{b]}^{~f} \chi_f ~R ~.
\end{eqnarray}
Using the constraint as appilied in the Jordan frame (\ref{constraintj}) and simplifying the above result, we have then for (\ref{waldonkilling}),
\begin{equation}
\chi_{[a} \nabla_{b]} \kappa ~\overset{\mathcal{H}^{(K)}}{=}~ - \frac{1}{ \phi} \Big(8 \pi \chi_{[a} T^{(m)f}_{b]} \chi_f + \chi^f (\chi_{[a}\nabla_{b]}\nabla_f ~\phi)\Big) ~.
\label{zeroth1j}
\end{equation}
Next, we have the fact that,
\begin{equation}
\pounds_{\boldsymbol{\chi}} (\nabla_a \phi) ~\overset{(\mathcal{M}, g_{ab}, \phi)}{=}~ 0 ~.
\label{lievarnablaphi}
\end{equation}
This is again to be expected since the scalar field $\phi$ is Lie transperted along $\boldsymbol{\chi}$ as evident under the constraint (\ref{constraintj}) and therefore the quantity $(\nabla_a \phi)$ is expected to satisfy the symmetry of the spacetime. However we give a brief sketch of this in Appendix \ref{AppFrob}. Using (\ref{lievarnablaphi}), we can verify that,
\begin{equation}
\chi^f (\chi_{[a}\nabla_{b]}\nabla_f ~\phi) ~\overset{\mathcal{H}^{(K)}}{=}~ 0 ~.
\label{frobenius}
\end{equation}
A detailed outlined proof of this is given in Appendix \ref{AppFrob}. So finally, we obtain from (\ref{zeroth1j}) and (\ref{frobenius}),
\begin{equation}
\chi_{[a} \nabla_{b]} \kappa ~\overset{\mathcal{H}^{(K)}}{=}~ - \frac{1}{ \phi} 8 \pi \chi_{[a} T^{(m)f}_{b]} \chi_f ~.
\label{zeroth2j}
\end{equation}
Next, we proceed to calculate $T^{(m)}_{~ab} \chi^a \chi^b$ on the Killing horizon. Using the field equations of motion (\ref{feomgabj}), we can show that,
\begin{eqnarray}
E_{ab} \chi^a \chi^b & ~\overset{(\mathcal{M}, g_{ab}, \phi)}{=}~ \frac{1}{16 \pi} \Big[\phi G_{ab} \chi^a \chi^b + \frac{\omega}{2 \phi} \boldsymbol{\chi}^2 ~(\nabla_i \phi \nabla^i \phi) - \frac{\omega}{\phi} (\chi^a \nabla_a \phi)~(\chi^b \nabla_b \phi) \nonumber
\\
& + \frac{1}{2}\boldsymbol{\chi}^2 V(\phi) - \chi^a \chi^b \nabla_a \nabla_b \phi  + \boldsymbol{\chi}^2 ~ (\nabla_i \nabla^i \phi)\Big] = \frac{1}{2} T^{(m)}_{~ab} \chi^a \chi^b ~.
\label{Tabchiachiab1j}
\end{eqnarray}
On the Killing horizon $\mathcal{H}^{(K)}$, $\boldsymbol{\chi}$ is null and the projection component $R_{ab}\chi^a \chi^b$ vanishes (\ref{Rabchiachibj}). Upon using the constraint relation (\ref{constraintj}), we obtain from (\ref{Tabchiachiab1j}), 
\begin{equation}
- \chi^a \chi^b \nabla_a \nabla_b \phi~\overset{\mathcal{H}^{(K)}}{=}~ 8 \pi T^{(m)}_{~ab} \chi^a \chi^b ~.
\label{Tabchiachib2j}
\end{equation}
Using the relation (\ref{lievarnablaphi}), it can be easily shown that $\chi^a \chi^b \nabla_a \nabla_b \phi$ vanishes on $\mathcal{H}^{(K)}$,
\begin{eqnarray}
\chi^a \chi^b \nabla_a \nabla_b \phi = \chi^a \Big(\pounds_{\boldsymbol{\chi}}(\nabla_a \phi) - \nabla_b \phi \nabla_a \chi^b \Big) ~\overset{\mathcal{H}^{(K)}}{=}~ - \kappa \chi^b \nabla_b \phi ~\overset{\mathcal{H}^{(K)}}{=}~ 0 ~.
\end{eqnarray}
Hence this allows us to finally conclude that,
\begin{equation}
T^{(m)}_{ab} \chi^a \chi^b ~\overset{\mathcal{H}^{(K)}}{=}~ 0 ~.
\label{Tabchiachibj}
\end{equation}
The above relation relation implies as usual that the vector field $T^{(m)a}_{~b} \chi^b$ lies on the tangent space of the Killing horizon $\mathcal{H}^{(K)}$ and hence is either null or spacelike. From the invariance of the matter action under conformal transformations,
\begin{equation}
\tilde{\mathcal{A}}^{(m)} = \int d^4 x \sqrt{-\tilde{g}} \tilde{\mathcal{L}}^{(m)} = \int d^4 x \sqrt{-g} \mathcal{L}^{(m)} = \mathcal{A}^{(m)} ~, 
\end{equation}
we necessarily have the following relation between the matter (and non-gravitational) Lagrangians  between the Einstein and the Jordan frames, under the conformal transformation rule (\ref{gabconformal}),
\begin{equation}
\tilde{\mathcal{L}}^{(m)} = \Omega^{-4} \mathcal{L}^{(m)} ~.
\end{equation}
From the defintion of the matter energy momentum tensor,
\begin{equation}
\tilde{T}^{(m)}_{ab} = \frac{2}{\sqrt{-\tilde{g}}} \frac{\delta}{\delta \tilde{g}^{ab}} ~ \Big(\sqrt{-\tilde{g}} ~\tilde{\mathcal{L}}^{(m)}\Big) = \Omega^{-4} \frac{\partial g^{cd}}{\partial \tilde{g}^{ab}} \frac{2}{\sqrt{-{g}}}\frac{\delta}{\delta g^{cd}} \Big(\sqrt{-g} ~ \mathcal{L}^{(m)}\Big) ~,
\end{equation}
we have the follwing relations between the matter energy momentum tensors in the two conformal frames,
\begin{equation}
\tilde{T}^{(m)}_{~ab} = \Omega^{-2} T^{(m)}_{~ab},~~~~~~~ \tilde{T}^{(m)a}_{~~b} = \Omega^{-4} T^{(m)a}_{~~b},~~~~~~~\tilde{T}^{(m)ab} = \Omega^{-6} T^{(m)ab}
\label{conformtab}
\end{equation}
Now, since $\Omega^2 = \phi$ is a strictly positive function of the spacetime coordinates, we conclude via (\ref{conformtab}) that if the NDEC holds in the Einstein frame, then it must also necessarily hold in the Jordan frame. The vector field $W^a$ defined as,
\begin{equation}
W^a \equiv -T^{(m)a}_{~~b}\chi^b ~,
\label{NDECj}
\end{equation}
is  future directed timelike or null for any future directed null vector field $\boldsymbol{\chi}$. But as again, we have previously shown that $T^{(m)a}_{~~b}$ can either be null or spacelike. Hence the NDEC as applied to $\mathcal{H}^{(K)}$ forces $T^{(m)a}_{~~b}$ to be null on the Killing horizon and hence is collinear to its null generators,
\begin{equation}
-T^{(m)a}_{~~b} \chi^b ~\overset{\mathcal{H}^{(K)}}{=}~ \alpha \chi^a ~, 
\end{equation}
where $\alpha$ is some proportionality factor. Finally using the above relation in (\ref{zeroth2j}), we get to our desired goal,
\begin{equation}
\chi_{[a} \nabla_{b]} \kappa ~\overset{\mathcal{H}^{(K)}}{=}~ 0 ~.
\label{zerothlawj}
\end{equation}
So we have essentially established the constancy of the surface gravity $\kappa$ over the Killing horizon $\mathcal{H}^{(K)}$ i.e the zeroth law holds for the Killing horizon established in the Jordan frame under the constraint (\ref{constraintj}).

\subsection{Approach II}
Now we give a different proof of the zeroth law in the two frames considered. However, this proof also relies upon the dynamical content of the theory in the sense that the field equations are used under the fact that the NDEC holds in both the frames. The method we follow is adopted from \cite{Gourg}. Let us begin with very generic considerations in the sense that suppose our spacetime $(\mathcal{M},g_{ab})$ admits a Killing vector field $\boldsymbol{\chi}$. The vector field $\boldsymbol{\chi}$ then generates the Killing horizon in the given spacetime, in the sense that $\boldsymbol{\chi}$ coincides with the null generators of the Killing horizon. The surface gravity $\kappa$ of the Killing horizon $\mathcal{H}$ is defined as,
\begin{equation}
\kappa^2 ~\overset{\mathcal{H}}{=}~ -\frac{1}{2} \nabla_a \chi_b \nabla^a \chi^b ~.
\label{surfacegrav}
\end{equation}
We can show, without using the gravitational field equations, that $\kappa$ is constant along the null generators. Afterall, this is to be expected since $\boldsymbol{\chi}$ is the symmetry generator of the horizon $\mathcal{H}$.
Taking directional derivative of the above equation (\ref{surfacegrav}) along the null generators $\boldsymbol{\chi}$, we have,
\begin{equation}
2 \kappa (\chi^i \nabla_i \kappa) ~\overset{\mathcal{H}}{=}~-(\nabla^a \chi^b) (\chi^i \nabla_i \nabla_a \chi_b) ~\overset{\mathcal{H}}{=}~ -(\nabla^a \chi^b) R_{baid} \chi^i \chi^d ~\overset{\mathcal{H}}{=}~ 0 ~.
\end{equation}
Since $\kappa$ is non-zero on the horizon (non-degenerate), we necessarily have,
\begin{equation}
(\chi^i \nabla_i \kappa) ~\overset{\mathcal{H}}{=}~ 0 ~.
\end{equation}
As a result once we have established the fact that we have respective Killing horizons in the two frames, we should be content in proving the constancy of the surface gravity only along the spacelike directions. This is exactly the point where we will require the respective field equations in the two frames. As before, we assume the Killing horizon $\mathcal{H}$ has a topology of $\mathbb{R}\times \mathcal{J}$, where $\mathcal{J}$ is a spacelike closed manifold transverse to the null generators. We can establish the relation \cite{Gourg},
\begin{equation}
{D}_c \kappa ~\overset{\mathcal{H}}{=} -R_{ab} \chi^a q^b_{~c} ~,
\end{equation}
where ${D}_c$ denotes the derivative w.r.t the spacelike manifold $(\mathcal{J},\boldsymbol{q})$ and $q^a_{~b} = \delta^a_{~b} + \chi^a k_b + k^a \chi_b$ denotes the induced metric on $\mathcal{J}$ with $k^a$ being the auxilairy null vector transverse to $\mathcal{H}$. 

\subsubsection{Einstein frame}
We now follow up with this in the Einstein frame $(\mathcal{M},\tilde{g}_{ab}, \tilde{\phi})$ where we have for the Killing horizon $\tilde{\mathcal{H}}^{(K)}$ generated by $\tilde{\boldsymbol{\chi}}$ (having the spacelike cross-section $(\tilde{\mathcal{J}}, \tilde{\boldsymbol{q}})$),
\begin{equation}
 \tilde{{D}}_c \tilde{\kappa} ~\overset{\tilde{\mathcal{H}}^{(K)}}{=}~ - \tilde{R}_{ab} \tilde{\chi}^a \tilde{q}^b_{~~c} ~.
\label{zerothgour1e}
\end{equation}
Using the field equations in the Einstein frame (\ref{feomgabe}) its quite easy to show that,
\begin{equation}
\tilde{R}_{ab} \tilde{\chi}^a \tilde{q}^b_{~c} = 16 \pi \Big(\frac{1}{2} \tilde{T}^{(m)}_{ab} \tilde{\chi}^a \tilde{q}^b_{c} + \frac{1}{2} (\tilde{\chi}^a \tilde{\nabla}_{a} \tilde{\phi}) ~\tilde{{D}}_c \tilde{\phi}\Big) ~.
\end{equation}
Use of the constraint relation (\ref{constrainte}) allows us to have,
\begin{equation}
\tilde{{D}}_c \tilde{\kappa} ~\overset{\tilde{\mathcal{H}}^{(K)}}{=}~ - 8 \pi \tilde{T}^{(m)}_{~ab} \tilde{\chi}^a \tilde{q}^b_{~c} ~.
\label{zerothgour2e}
\end{equation}
Invoking the validity of the NDEC as applied to the Einstein frame, we have,
\begin{equation}
-\tilde{T}^{(m)}_{ab} \tilde{\chi}^a ~\overset{\tilde{\mathcal{H}}^{(K)}}{=}~  \tilde{\beta} \tilde{\chi}_b ~,
\end{equation}
where $\tilde{\beta}$ is some proportionality factor. This further allows us to conclude that the R.H.S of (\ref{zerothgour2e}) on the Killing horizon $\tilde{\mathcal{H}}^{(K)}$ is,
\begin{equation}
- 8 \pi \tilde{T}^{(m)}_{~ab} \tilde{\chi}^a \tilde{q}^b_{~c}  ~\overset{\tilde{\mathcal{H}}^{(K)}}{=}~ - 8 \pi \tilde{\beta} ~\tilde{\chi}_b \tilde{q}^{b}_{~c} = 0 ~. 
\end{equation}
The last part comes from the fact that the null generator of $\tilde{\mathcal{H}}^{(K)}$ is orthogonal to the spacelike cross section $\tilde{\mathcal{J}}$. So in essense, we have finally showed that in the Einstein frame, the zeroth law holds, 
\begin{equation}
\tilde{{D}}_c \tilde{\kappa} ~\overset{\tilde{\mathcal{H}}^{(K)}}{=}~ 0~.
\label{zerothgoure}
\end{equation} 

\subsubsection{Jordan frame}
We now proceed towards the Jordan frame where we have the relation established on the Killing horizon $\mathcal{H}^{(K)}$ (with the spacelike cross-section $(\mathcal{J}, \boldsymbol{q})$),
\begin{equation}
 {D}_c \kappa ~\overset{\mathcal{H}^{(K)}}{=}~ -R_{ab} \chi^a q^b_{~c} ~.
\label{zerothgour1j}
\end{equation}
Again, using the field equations (\ref{feomgabj}) for the Jordan frame and the constraint (\ref{constraintj}) it is quite easy to show that,
\begin{equation}
R_{ab} \chi^a q^b_{~c} = \frac{1}{\phi} \Big(8 \pi T^{(m)}_{ab} \chi^a q^b_{~c} + \chi^a q^b_{~c} \nabla_a \nabla_b \phi\Big) ~.
\end{equation}
The quantity $\chi^a q^b_{~c} \nabla_a \nabla_b \phi$ vanishes on the Killing horizon $\mathcal{H}^{(K)}$,
\begin{equation}
\chi^a q^b_{~c} \nabla_a \nabla_b \phi  ~\overset{\mathcal{H}^{(K)}}{=}~ 0 ~.
\label{frobe1j}
\end{equation} 
This has has been shown in Appendix \ref{AppFrob1}. This allows us again to have,
\begin{equation}
 {D}_c \kappa ~\overset{\mathcal{H}^{(K)}}{=}~ - \frac{1}{\phi} 8 \pi T^{(m)}_{ab} \chi^a q^b_{~c} ~.
\label{zerothgour2j}
\end{equation}
Similar validity of the NDEC in the Jordan frame allows us to establish the fact that the R.H.S of (\ref{zerothgour2j}) vanishes on $\mathcal{H}^{(K)}$. Hence we finally establish the zeroth law as well in the Jordan frame.
\begin{equation}
 {D}_c \kappa ~\overset{\mathcal{H}^{(K)}}{=}~ 0~.
\label{zerothgourj}
\end{equation} 
Since temperature is proportional to surface gravity, the above analysis shows that the temperaure is constant over the horizon. This we have shown separately in both the frames. We 
\section{Conclusion}
There has been much debate about the physical (in)equivalence of the Jordan and the Einstein frame and the question still remains as to what can be considered ``more'' physical than the other. Any establishment of (in)equivalences of physical and thermodynamical quantities can only help us to address such long-standing issues. Our present work has been focussed in this particular direction aimed at the thermodynamical aspects of the gravitational theories in the two frames. In the earlier works, it had been shown in the context of Killing horizons present in the spacetime that the thermodynamic parameters are equivalent in the two frames. However, the presence of the Killing horizon imposes symmetry requirements on the spacetime. Moreover, since the Killing horizon describes a stationary equilibrium black hole system, the equivalence of such thermodynamic parameters is restricted only to equilibrium processes. However, it has been established that at least for Einstein gravity and the Lanczos-Lovelock gravity the gravitational field equations expressed in the neighbourhood of a generic null hypersurface asssumes a thermodynamic interpretation in analogy with the first law of thermodynamics. The presence of this generic null surface does not ask for any symmetry requirements on the spacetime. The resulting thermodynamical interpretation given under the context of virtual displacement of the null hypersurface incorporates both equilibrium as well as non-equilibrium processes. That is such an interpretation is capable of handing internal entropy generation due to dissipation or viscous effects under the process of virtual displacement \cite{Dey:2020tkj}. We have shown that such an equivalence of the relevant thermodynamic parameters also exists in the case of ST theory of gravity. For this, we used the projection component $R_{ab} l^a k^b$ onto a generic null hypersurface established in both the Einstein and the Jordan frames. Through their respective dynamics and the process of virtual displacements, we connected the component $R_{ab}l^a k^b$ to the relevant thermodynamic identities (established on the null hypersurface) in both the frames in a completely covariant fashion. We stress again that such an identity has been interpreted not via any coordinate system adapted to the null hypersurface (say the Gaussian null coordinate system). Our analysis has been done completely in a covariant fashion which allows us to provide covariant expressions of the relevant thermodynamical parameters, which can then be adapted to any coordinate system of the person's choice describing the spacetime in the neighbourhood of the null surface. This allowed us to interpret from the analogical thermodynamical first law established in both the frames, that the quantities like temperature, entropy density, energy and the work function are equivalent in both the frames. Finally, this nicely ties in with another interpretaion provided under the umbrella of the projection component $R_{ab} l^a q^b_{~c}$ which leads to the Damour-Navier-Stokes equation. The equivalence of the relevant fluid variables (of the DNS equation) in the two frames had previously been estblished. Thus such fluid variables and thermodynamic parameters operate on an equal footing when the two frames are considered. This we hope lends much ground to the issue about the physical equivalences between the two frames.

Let us reinstate the fact that the above thermodynamical interpretation (using the field equations in the two frames) had been drawn based on analogy with conventional thermodynamics. This however does not allow concrete physical interpretation of the thermodynamic variables, especially the temperature. In conventional thermodynamics, the temperature is essentially an intensive variable whose constancy defines the notion of thermal equilibrium between two thermal systems under contact. This is essentially the statement of the zeroth law of thermodynamics. However, for gravitational dynamics, there is actually no notion of two black hole systems being in thermal equilibrium with each other. The zeroth law of black hole mechanics says that a black hole system in thermal equilibrium must be by definition a Killing horizon (defining a stationary black hole system) over which its surface gravity is constant. This constancy of the surface gravity allows us then to give a concrete identification and interpretation of the temperature associated with any generic null hypersurface. So our analysis would be quite well rounded if we could prove the zeroth law as established in Killing horizons in the two frames. The zeroth law for Scalar-Tensor theory had been established in the literature under the constraint that the scalar field needed to constant over the Killing horizon. However, we believe that this is a bit too restrictive. In second part of our analysis, we showed that in order for the zeroth law to hold in both the frames, the only requirement we demand of the scalar field is for it to respect the symmetry of the given spacetime. That is, we only demanded that the scalar field is Lie transported along the symmetry generator of the spacetime. This implies that on the  Killing horizon the scalar field is independant of the coordinate along the null symmetry generator, but can very well depend on the angular/transverse coordinates. We did not put any extra symmetries on the spacetime other than to impose that the matter and non-gravitational fields in the two frames satisfy the NDEC. 

Finally, we believe that our results based on the thermodynamic identity valid on any generic null hypersurface and the proof of the zeroth law in the two frames provide some clarifications into questions regarding their physical equivalences (or inequivalences for that matter). It is worthwhile to mention that at the classical level a certain class of $f(R)$ gravity can be cast in the form ST theories (as not possible in general; for instance see \cite{Castaneda:2018cxx} and references therein). The thermodynamic structure of such $f(R)$ theories can be discussed along the present line of thought.  We hope that our analysis will help shed more light onto the nature of physics in both the Einstein and the Jordan frame.

\vskip 3mm
\noindent
{\bf Acknowledgement:} The research of one of the authors (B.R.Majhi) is supported by Science and Engineering Research Board (SERB), Department of Science \& Technology (DST),  Government of India, under the scheme Core Research Grant (File no. CRG/2020/000616).

\vskip 5mm
\noindent
\appendix{\bf{Appendices}}

	\section{Equivalence of Eq. \eqref{RELTN} and Eq. \eqref{RELTNJOR} under the conformal transformation}\label{AppA}
	Using eq. \eqref{TRANSREL}, one obtains the following relations
	\begin{align}
	\t D_a\t\O^a= \frac{1}{\phi} D_a\O^a+\frac{1}{2\phi}\Big[\theta_{(l)}k^i+\theta_{(k)}l^i\Big]\na_i(\ln\phi)+\frac{q^{ab}}{2\phi}\na_a\na_b(\ln\phi)~. \label{TRAN5}
	\end{align}
	
	\begin{align}
	\t\theta_{(\t{l})}\t\theta_{(\t{k})}=\frac{1}{\phi}\Big[\theta_{(l)}\theta_{(k)}+\Big(\theta_{(l)}k^i+\theta_{(k)}l^i\Big)\na_i(\ln\phi)+l^ik^j\na_i(\ln\phi)\na_j(\ln\phi)]~.\label{TRAN6}
	\end{align}
	
	\begin{align}
	\t l^i\t\na_i\t\theta_{(\t{k})}=\frac{1}{\phi}l^i\na_i\theta_{(k)}+\frac{1}{\phi}\Big[\O^i-\kappa k^i-\theta_{(k)}l^i\Big]\na_i(\ln\phi)-\frac{l^ik^j}{\phi}\na_i(\ln\phi)\na_j(\ln\phi)+\frac{l^ik^j}{\phi}\na_i\na_j(\ln\phi)~.\label{TRAN7}
	\end{align}
	Using \eqref{TRAN1}, \eqref{TRAN2}, \eqref{TRAN3}, \eqref{TRAN5}, \eqref{TRAN6} and \eqref{TRAN7} in \eqref{RELTN} one obtains \eqref{RELTNJOR}.

	\section{Obtaining Eq. \eqref{RELATHER} starting from Eq. \eqref{RELTNJOR}}\label{AppB}
	
	We start from \eqref{RELTNJOR} and write each parameters of the Jordan frame (such as $\theta_{(l)}$, $\theta_{(k)}$, $\kappa$, $\Omega^a$ \textit{etc.}), in terms of the parameters of the Einstein frame (such as $\t\theta_{(\t{l})}$, $\t\theta_{(\t{k})}$, $\t\kappa$, $\t\Omega^a$ \textit{etc.}) using Eq. \eqref{TRANSREL}. We consider term-by-term of Eq. \eqref{RELTNJOR} and obtain the following relations for each term
	\begin{align}
	\kappa\theta_{(k)}=\phi\t\kappa\t\theta_{(\t{k})}-\t\kappa k^i\na_i(\ln\phi)-\phi\t\theta_{(\t{k})}l^i\na_i(\ln\phi)+l^ik^j\na_i(\ln\phi)\na_j(\ln\phi)~. \label{T11}
	\end{align}

	\begin{align}
	D_a\O^a=\phi D_a\t\O^a+\t\O^a D_a\phi-\frac{1}{2}D^aD_a(\ln\phi)~.\label{T22}
	\end{align}
	
	\begin{align}
	\theta_{(l)}\theta_{(k)}=\phi\t\theta_{(\t{l})}\t\theta_{(\t{k})}-\phi\t\theta_{(\t{k})}l^i\na_i(\ln\phi)-\t\theta_{(\t{l})}k^i\na_i(\ln\phi)+l^ik^j\na_i(\ln\phi)\na_j(\ln\phi)~.\label{THETALTHETAK}
	\end{align}
	Now, we know that $D^aD_a(\ln\phi)=q^i_j\na_i(q^{jk}\na_k(\ln\phi))$, from which it can be obtained that
	\begin{align}
	D^aD_a(\ln\phi)=q^{ij}\na_i\na_j(\ln\phi)+[\theta_{(\t{l})}k^i+\theta_{(\t{k})}l^i]\na_i(\ln\phi)~.
	\end{align}
	Writing $\theta_{(l)}$ and $\theta_{(k)}$ in terms of $\t\theta_{(\t{l})}$ and $\t\theta_{(\t{k})}$, one further obtains 
	\begin{align}
	D^aD_a(\ln\phi)=q^{ij}\na_i\na_j(\ln\phi)+\phi\t\theta_{(\t{k})}l^i\na_i(\ln\phi)+\t\theta_{(\t{l})}k^i\na_i(\ln\phi)-2l^ik^j\na_i(\ln\phi)\na_j(\ln\phi)~. \label{TEMP1}
	\end{align}
	Using \eqref{TEMP1} in \eqref{THETALTHETAK}, one obtains
	\begin{align}
	\theta_{(l)}\theta_{(k)}=\phi\t\theta_{(\t{l})}\t\theta_{(\t{k})}-D^aD_a(\ln\phi)+q^{ij}\na_i\na_j(\ln\phi)-l^ik^j\na_i(\ln\phi)\na_j(\ln\phi)~.\label{T33}
	\end{align}
	Writing $\O^a$ in terms of $\t\O^a$ we obtain
	\begin{align}
	\O^a\O_a=\phi\t\O^a\t\O_a-\phi\t\O^i\na_i(\ln\phi)+\frac{1}{4}q^{ij}\na_i(\ln\phi)\na_j(\ln\phi)~.\label{T44}
	\end{align}
	Finally we obtain
	\begin{align}
	l^i\na_i\theta_{(k)}=\phi l^i\na_i\t\theta_{(\t{k})}+\phi\t\theta_{(\t{k} )}l^i\na_i(\ln\phi)-\phi\t\O^i\na_i(\ln\phi)+\frac{1}{2}q^{ij}\na_i(\ln\phi)\na_j(\ln\phi)+\phi\t\k\t k^i\na_i(\ln\phi)
	\nonumber
	\\
	-l^ik^j\na_i(\ln\phi)\na_j(\ln\phi)-l^ik^j\na_i\na_j(\ln\phi)~.\label{T55}
	\end{align}
	Now, using \eqref{T11}, \eqref{T22}, \eqref{T33}, \eqref{T44} and \eqref{T55} in Eq. \eqref{RELTNJOR}, one obtains Eq. \eqref{RELATHER}.
	\section{Proof the zeroth law for static spacetimes}\label{staticzeroth}
	We assume the spacetime $(\mathcal{M}, \boldsymbol{g})$ to be static i.e both stationary (admitting a killing vector field $\boldsymbol{\chi}$) and the Killing vector field $\boldsymbol{\chi}$ as being hypersurface orthogonal. The hypersurface orthogonality condition implies that over the manifold we have,
	\begin{equation}
	\chi_{[c}\nabla_{b} \chi_{a]} = \chi_c \nabla_b \chi_a + \chi_b \nabla_a \chi_c + \chi_a \nabla_c \chi_b = 0 ~.
	\end{equation}
The above relation is valid on the manifold and not just only on the Killing horizon $\mathcal{H}^{(K)}$ in the given spacetime. Hence we can safely take the derivative of the above relation:
\begin{equation}
\nabla^a \Big[\chi_c \nabla_b \chi_a + \chi_b \nabla_a \chi_c + \chi_a \nabla_c \chi_b\Big] = 0 ~.
\end{equation}
Using the Killing equation $\nabla_a \chi_b + \nabla_b \chi_a = 0$, its quite easy to show that the above relation reduces to,
\begin{equation}
\chi_c \nabla_a \nabla_b \chi^a + \chi_b \square \chi_c  + \chi^a \nabla_a \nabla_c \chi_b  = 0 ~.
\label{stat1}
\end{equation}
We now define the following quantity $P_a$ to be,
\begin{equation}
P_a = R^f _{~a} \chi_f ~.
\end{equation}
Simple manipulations allow us to have,
\begin{eqnarray}
P_a = R^f _{~a} \chi_f = [\nabla_b,\nabla_a]\chi^b& \nonumber \\
=\nabla_b \nabla_a \chi^b - \nabla_a \nabla_b \chi^b = \nabla_b \nabla_a \chi^b  = -\nabla_b \nabla^b \chi_a = -\square \chi_a& ~.
\label{Pviabox}
\end{eqnarray}
Use of Eq. (\ref{Pviabox}) in Eq. (\ref{stat1}) yields,
\begin{equation}
\chi_c P_b - \chi_b P_c =  - \chi^a \nabla_a \nabla_c \chi_b ~.
\label{E6}
\end{equation}
Since $\boldsymbol{\chi}$ is a symmetry generator of the spacetime, any tensor constructed out of $\boldsymbol{\chi}$ and $g_{ab}$ will also respect the spacetime symmetry. Such a tensor field is the quantity $T_{cb} = \nabla_c \chi_b$. Explicitly, this means that,
\begin{equation}
\pounds_{\boldsymbol{\chi}} T_{cb} = 0 ~.
\end{equation}
This follows that,
\begin{eqnarray}
&\chi^a \nabla_a \nabla_c \chi_b + \nabla_a \chi_b  (\nabla_c \chi^a) + \nabla_c \chi_a (\nabla_b \chi^a) = 0\nonumber \\
&\chi^a \nabla_a \nabla_c \chi_b = 0 ~.
\label{E8}
\end{eqnarray}
Use of Eq. (\ref{E8}) in Eq. (\ref{E6}) implies,
\begin{equation}
\chi_c P_b - \chi_b P_c = 0 ~.
\label{E9}
\end{equation}
Now, as applied onto the Killing horizon $\mathcal{H}^{(K)}$, we have the relation (\ref{waldonkilling}), upon which using Eq. (\ref{E9}) leads to,
\begin{eqnarray}
\chi_d~ \chi_{[a} \nabla_{b]} \kappa &\overset{\mathcal{H}^{(K)}}{=}& - \chi_d \chi_f R^f_{~[b} \chi_{a]} \nonumber \\
&\overset{\mathcal{H}^{(K)}}{=}& - \frac{\chi_d}{2} \Big(\chi_f R^f_{~b} \chi_a - \chi_f R^f_{~a} \chi_b\Big) \nonumber \\
&\overset{\mathcal{H}^{(K)}}{=}& - \frac{\chi_d}{2}(P_b \chi_a - P_a \chi_b) = 0 ~.
\end{eqnarray}
Since we assume our Killing horizon to be non-degenerate, we essentially have $\chi_{[a} \nabla_{b]}\kappa \overset{\mathcal{H}^{(K)}}{=} 0$. This essentially proves the zeroth law for static spacetimes admitting a Killing horzion without the need of the dynamical field equations.
	\section{Proof of the relation (\ref{frobenius})}\label{AppFrob}
	We begin by showing that the Lie derivative of $\nabla_a \phi$ along the null generator of $\boldsymbol{\chi}$ vanishes over $(\mathcal{M}, g_{ab}, \phi)$ using the constraint (\ref{constraintj}), 
	\begin{align}
	\pounds_{\boldsymbol{\chi}} (\nabla_a \phi) =& \chi^f \nabla_a \nabla_f \phi + (\nabla_f \phi) (\nabla_a \chi^f) \nonumber 
	\\
	= &\nabla_{a}(\chi^f \nabla_f \phi) - (\nabla_a \chi^f)(\nabla_f \phi) + (\nabla_f \phi)(\nabla_a \chi^f) = 0 ~.
	\label{liej}
	\end{align}
	We have then,
	\begin{equation}
	\chi^f (\chi_{[a}\nabla_{b]}\nabla_f ~\phi) = \frac{1}{2} \Big(\chi^f\chi_a \nabla_b \nabla_f \phi - \chi^f \chi_b \nabla_a \nabla_f \phi \Big) ~.
	\end{equation}
	Using the first line of (\ref{liej}) we have $\chi_b \chi^f \nabla_f \nabla_a \phi = - \chi_b (\nabla_c \phi)(\nabla_a \chi^c)$. Putting this in the second term of above relation, we have,
	\begin{align}
\chi^f (\chi_{[a}\nabla_{b]}\nabla_f ~\phi) =& \frac{1}{2} \Big(\chi_a \nabla_b (\chi^f \nabla_f \phi) - \chi_a (\nabla_b \chi^f) (\nabla_f \phi) + \chi_b (\nabla_f \phi)(\nabla_a \chi^f)\Big) \nonumber 
	\\
	=&\frac{1}{2} \nabla^f \phi \Big(\chi_a \nabla_f \chi_b + \chi_b \nabla_a \chi_f\Big) ~.
	\end{align}
	Let us then invoke the hypersurface orthogonality of the integrable null hypersurface $\mathcal{H}^{(K)}$ generated by the null vector field $\boldsymbol{\chi}$ in the absense of torsion,
	\begin{equation}
	\chi_a \nabla_f \chi_b +\chi_f \nabla_b \chi_a + \chi_b \nabla_a \chi_f  ~\overset{\mathcal{H}^{(K)}}{=}~ 0 ~.
	\label{hypoth}
	\end{equation} 
	Using the relation (\ref{hypoth}), we have,
	\begin{equation}
	\chi^f (\chi_{[a}\nabla_{b]}\nabla_f ~\phi) ~\overset{\mathcal{H}^{(K)}}{=}~ -\frac{1}{2} \Big(\chi^f \nabla_f \phi (\nabla_b \chi_a)\Big) ~\overset{\mathcal{H}^{(K)}}{=}~ 0 ~.
	\end{equation}
	This proves our desired relation.
	
	\section{Proof of the relation (\ref{frobe1j}) }\label{AppFrob1}
	
	Next we proceed to give a proof of (\ref{frobe1j}). Using the relation (\ref{liej}), we have on the Killing horizon $\mathcal{H}^{(K)}$,
	\begin{align}
	\chi^a q^b_{~c} \nabla_a \nabla_b \phi =& -q^b_{~c} (\nabla_a \phi)(\nabla_b \chi^a) = -(\delta^b_{~c} + \chi^b k_c + k^b \chi_c) (\nabla_a \phi) (\nabla_b \chi^a) \nonumber 
	\\
	=& - \nabla^a \phi \Big(\nabla_c \chi_a + \kappa k_c \chi_a + k^b \chi_c \nabla_b \chi_a \Big) ~.
	\end{align}
	From the hypersurface orthogonality condition of the Killing horizon (\ref{hypoth}), we have,
	\begin{align}
	\chi^a q^b_{~c} \nabla_a \nabla_b \phi &~\overset{\mathcal{H}^{(K)}}{=}~ -\nabla^a \phi \Big(\nabla_c \chi_a + \kappa k_c \chi_a - k_b (\chi_b \nabla_a \chi_c + \chi_a \nabla_c \chi_b)\Big) \nonumber
	\\
	&~\overset{\mathcal{H}^{(K)}}{=}~ -\nabla^a \phi \Big(\nabla_c \chi_a + \kappa k_c \chi_a + \nabla_a \chi_c - (k^b \nabla_c \chi_b) \chi_a\Big) ~.
	\end{align}
	Use of the fact that $\boldsymbol{\chi}$ is a symmetry generator of $\mathcal{H}^{(K)}$ and the constraint condition (\ref{constraintj}), allows us to have,
	\begin{equation}
	\chi^a q^b_{~c} \nabla_a \nabla_b \phi ~\overset{\mathcal{H}^{(K)}}{=}~ 0 ~.
	\end{equation}

\end{document}